\documentclass[12pt,nofootinbib]{revtex4}
\usepackage{amssymb}
\usepackage{amsmath}
\usepackage{epsfig}
\usepackage{tabularx}
\usepackage{inputenc}
\usepackage{url}
\usepackage{breqn}
\usepackage{hyperref}
\usepackage{footmisc}
\usepackage{titlesec}
\usepackage{fancyhdr}
\usepackage{graphicx}
\usepackage{caption}
\usepackage{wrapfig}
\graphicspath{{latex_images/}}

\def\tanb{\tan\beta}
\def\no{\tilde{\chi}^0_1}

\def\simlt{\stackrel{<}{{}_\sim}}
\def\simgt{\stackrel{>}{{}_\sim}}

\titleformat{\section}
  {\large\bfseries\scshape}{\thesection}{1em}{}

\begin{document}
\begin{titlepage}
\begin{center}
\hfill EFI-17-1
\end{center}
\title{Constraints on Supersymmetric Dark Matter for Heavy Scalar Superpartners}
\vspace{1.0cm}
\author{\textbf{Peisi Huang$^{a,d,e}$, Roger A. Roglans$^{c}$, Daniel D. Spiegel$^{c}$ \\ Yitian Sun$^{c}$,  and Carlos E.M. Wagner$^{a,b,d}$} \\
\vspace{1.0cm}
\normalsize\emph{$^a$Enrico Fermi Institute $\&$ $^b$Kavli Institute for Cosmological Physics,}\\
\normalsize\emph{$^c$University of Chicago, Chicago, IL 60637} \\
\normalsize\emph{$^d$HEP Division, Argonne National Laboratory, 9700 Cass Ave., Argonne, IL 60439} \\
\normalsize\emph{$^e$Physics  and Astronomy Department, Texas A$\&$M University, 
College Station, TX 77843}
\vspace{1.5cm}
}

\begin{abstract}
We study the constraints on neutralino dark matter in minimal low energy supersymmetry models and the case of heavy lepton 
and quark scalar superpartners.  For values of the Higgsino and gaugino mass parameters of the order of the weak scale, direct 
detection experiments  are already  putting strong bounds on  models in which the dominant interactions
between the dark matter candidates and nuclei are governed by Higgs boson exchange processes, particularly for positive
values of the Higgsino mass parameter $\mu$.  For negative values of $\mu$, there can be destructive interference between 
the amplitudes associated with the exchange of the standard CP-even Higgs boson and the exchange of the non-standard one. This leads to 
specific regions of parameter space which are consistent with the current experimental constraints and a thermal origin of the observed relic density. 
In this article we  study the current experimental constraints on these scenarios, as well as  the  future experimental  probes, using a combination  of direct and indirect dark matter detection and heavy Higgs and electroweakino  searches at hadron colliders. 
	
\end{abstract}

\maketitle
\end{titlepage}
\section{Introduction}
Since its proposal in 1933 by Fritz Zwicky~\cite{zwicky}, the existence of dark matter (DM) has been supported by many indirect detection measurements. Besides observations of galaxy clusters and the magnitude of gravitational lensing, the rotation curves of spiral galaxies provide evidence that a significant portion of these galaxies is made of nonluminous matter~\cite{bosma,broeils}. Recently, the density of cold dark matter in the universe was estimated by the Planck collaboration to be  $\Omega_ch^2 = 0.1198 \pm 0.0015$~\cite{planck}. Overall, the evidence indicates that a significant portion of matter in the universe is non-baryonic, but the exact nature of dark matter is still unknown. One favored dark matter candidate is a weakly interacting massive particle (WIMP)\textemdash an uncharged, colorless, stable particle with a heavy enough mass to cease annihilation in the early stages of the universe, thus leaving behind the substantial cosmological abundance seen today~\cite{kamio}. No particles in the Standard Model (SM) account for all these properties, so we are forced to look at theories of physics beyond the SM (BSM).

Supersymmetry (SUSY) provides a well-motivated extension of the standard model that  contains such a WIMP in the form of the lightest supersymmetric particle (LSP)~\cite{Jungman:1995df}. In the Minimal Supersymmetric Standard Model (MSSM) with $R$-parity conservation, the LSP is typically the lightest neutralino $\widetilde \chi_1^0$.

Searches for BSM particles at the LHC have not been fruitful, and these searches put constraints on supersymmetric models. In certain models, the LHC searches now limit the mass of gluinos and the first two generations of squarks to be above 1.5 TeV~\cite{atlas,cms}. Other production analyses have constrained chargino and neutralino masses to be at least of the order of 100 GeV~\cite{atlasew,cmsew}, complementing the existing
bounds from LEP2 searches~(see, for example, Ref.~\cite{Abdallah:2003xe}). In addition, searches for heavy Higgs bosons constrain MSSM parameters such as $\tan \beta$ and $M_A$, the mass of the CP-odd Higgs boson~\cite{CMS:htautau2016}.

Direct dark matter detection experiments (DDMD), such as LUX~\cite{luxfinal}, PICO~\cite{pico}, XENON100~\cite{xenon100}, and PandaX~\cite{panda}, have also so far come up empty-handed. These experiments set upper bounds on both the spin-dependent (SD) and spin-independent (SI) cross sections of WIMPs scattering off nucleons. LUX presents the strongest bounds---their most recent data limits the WIMP-nucleon spin independent cross section $\sigma_{SI}$ to be lower than a few times $10^{-10}$ pb for a WIMP of mass in the range 20 GeV~$\simlt m_{\chi} \simlt 200$ GeV~\cite{luxfinal}. Future DDMD experiment such as LZ~\cite{lz} and Xenon1T~\cite{Xenon1T} will probe regions of SI cross sections
two orders of magnitude lower than those probed at present, and therefore it is interesting to explore the implications of the (non)observation 
of a signal in these experiments. 

Cheung et. al.~\cite{Cheung:BS} have identified regions of interest in the $M_1$ (bino mass), $M_2$ (wino mass), and $\mu$ (higgsino mass) parameter space called ``blind spots'', where the SI or SD cross sections are suppressed due to vanishing couplings of the LSP to the lightest CP-even Higgs and to the Z boson~\cite{Cheung:BS}. They worked in the decoupling regime, for very large values of  the heavy Higgs boson
masses, squarks, and sleptons such that they no longer affect properties relevant to DM.   For smaller values of the heavy Higgs boson masses,
and negative values of $\mu$, there is in general a destructive interference between the contributions to the SI cross section amplitude coming from the exchange of the standard CP-even Higgs boson and the non-standard one.  This leads to a cancellation in the total SI cross section amplitude in certain regions of parameters.  This effect was first noticed numerically, while performing a scan over the MSSM  parameter space~\cite{Mandic:2000jz,Ellis:2000ds,PhysRevD.63.065016,Ellis:2005mb,Baer:2006te,Han:2014nba,Baer:2015tva}.
In Ref.~\cite{Huang:2014xua}, an analytical expression for the relation between different parameters necessary to reach these
generalized blind spot scenarios was presented.  
The present DDMD constraints are
still relatively weak and  allow for a wide range of parameters, which can significantly deviate from the ones associated with the blind spot scenario. However,
as we will discuss in this article,  if future DDMD experiments continue to strengthen these constraints, it will become necessary to consider regions of parameter space close to the blind spot scenarios.

In this paper, we explore the current constraints on these scenarios, putting emphasis on regions of parameters consistent with  the observed thermal relic density, and checking these regions against DDMD, Higgs and BSM searches at the LHC. In section~\ref{sec:cal}, we discuss the theoretical basis and give an analytical formula for the generalized blind spot scenario. In section~\ref{sec:DDMDC}, we analyze the regions in the $\mu-M_1$ parameter space allowed by SI direct detection bounds and relic density considerations. In section~\ref{sec:LHC}, we test these regions against searches at the LHC for electroweakinos and heavy Higgs bosons. We also discuss the constraints coming from SD and indirect dark matter detection experiments, as well as from precision measurements of the observed standard model Higgs. We reserve section \ref{sec:conc} for our conclusions.

\section{DDMD and Blind spots}
\label{sec:cal}

In the MSSM, assuming heavy squarks and sleptons, the neutralino SI scattering process is mediated by the exchange of CP-even Higgs bosons. 
For Higgsino and gaugino mass parameters of the order of the weak scale the typical SI scattering cross section through the 125 GeV Higgs  boson
is of the order 10$^{-45}$ cm$^2$, which is in tension with the LUX results~\cite{luxfinal}. 
A possible way to suppress the SI scattering cross section is to decouple the heavy Higgs boson and suppress the lightest neutralino coupling to the 125~GeV Higgs boson. In the decoupling limit, the lightest and heaviest CP-even Higgs bosons in the MSSM, $h$ and $H$, are given by
\begin{eqnarray}
\sqrt{2} h & = & \cos\beta H_d^0 + \sin\beta H_u^0 
\nonumber\\  
\sqrt{2} H & = & \sin\beta H_d^0 - \cos\beta H_u^0
\end{eqnarray}
where $H_u^0$ and $H_d^0$ are the real components of the neutral Higgs bosons that couple to the up and down
quarks, respectively, and $\tan\beta = v_u/v_d$ is the ratio of the Higgs vacuum expectation values.  Given the neutralino mass
and  interactions terms,
\begin{eqnarray}
{\cal L}  & \supset & -\sqrt{2}g' Y_{H_{u}}\tilde{B}\tilde{H}_uH_u^{*}-\sqrt{2}g\tilde{W}^a\tilde{H}_ut^aH_u^{*}+(u \leftrightarrow d) + h.c.
\nonumber\\
& - &  \left( \frac{M_1}{2} \tilde{B} \tilde{B} +  \frac{M_2}{2} \tilde{W}^a \tilde{W}^a  + \mu \tilde{H}_d \tilde{H}_u + h.c. \right)
\label{eq:lag}
\end{eqnarray}
where $Y_i$ is the $H_i$ hypercharge, $\tilde{B}$ and $\tilde{W}$ are the superpartners of the neutral hypercharge (Bino) and
weak gauge bosons (Winos), $\tilde{H}_i$ are the superpartners of the Higgs bosons (Higgsinos), $M_{1,2}$ denote the Bino and neutral Wino
mass parameters and $\mu$ is the Higgsino mass parameter. One can show that the neutralino 
coupling to the lightest Higgs vanishes when
\begin{equation}
m_{\chi} + \mu \sin 2\beta = 0,
\end{equation}
where $m_{\chi}$ is the mass of the lightest neutralino.  Under the above conditions, for very
large values of the heavy Higgs boson masses, the tree-level contribution to the SI scattering cross section vanishes, which is identified as a blind spot in direct detection experiments~\cite{Cheung:BS}.   

Another way to suppress the SI scattering cross section, when the heavy Higgs is not too heavy, is to have destructive interference between the 125 GeV Higgs exchange and the heavy Higgs exchange amplitudes.  In this case, one goes away from the decoupling limit and the CP-even
Higgs mixing angle may no longer be identified with $\beta$. However, the deviations of the mixing angle from the decoupling values tend to be
small in the region of parameters of interest for this paper, and therefore, for simplicity we will keep the expressions valid close to the decoupling
limit. In such a case, the  amplitude of the scattering cross section of down-quarks to neutralinos is proportional to 
\begin{equation}
a_d \sim \frac{m_d}{\cos\beta} \left(\frac{\cos\beta \  g_{\chi \chi h}}{m_h^2}+\frac{\sin\beta  \  g_{\chi\chi H}}{m_H^2} \right),
\label{eq:sig_d}
\end{equation}
where $m_h$ and $m_H$ are the masses of the lightest and heavy CP-even Higgs bosons, respectively, and $g_{\chi\chi h}$  and
$g_{\chi\chi H}$ are their couplings to the lightest neutralino.
In general, the coupling of the neutralinos to the heavy and light Higgs bosons are similar in magnitude, and they may
differ in sign. The heavy Higgs contribution, although suppressed by the square of $m_H$ is 
enhanced by $\tan\beta$. Therefore, for moderate or large values of $\tan\beta$ the light and heavy Higgs contributions to the amplitude
may be similar in magnitude and may interfere destructively. 
 
Taking into account also the interaction of neutralinos with up-quarks in order to define the neutralino interaction with nuclei, one
can show that the SI scattering cross section is
proportional to~\cite{Huang:2014xua} 
\begin{equation}
\sigma_p^{SI} \sim \left[(F_{d}^{(p)}+F_{u}^{(p)})(m_{\chi}+\mu \sin2\beta)\frac{1}{m_h^2}+\mu \tan\beta \cos2\beta(-F_{d}^{(p)}+F_u^{(p)}/\tan^2\beta)\frac{1}{m_H^2}\right]^2,
\label{eq:sig}
\end{equation}
with $F_{u}^{(p)} \approx 0.15$ and $F_{d}^{(p)}  \approx 0.14$. 
The first term denotes the contribution of the lightest Higgs and its cancellation leads to the traditional blind spot scenarios
discussed above~\cite{Cheung:BS}.  The second term is  the contribution of the heavy Higgs and  as mentioned before for values of $|\mu|\,\simgt m_\chi$ and large $\tan\beta$ may become of the same order as the SM-like Higgs one.

For moderate or large values of $\tan\beta$, the tree-level contribution, mediated by the CP-even Higgs bosons,  vanishes when~\cite{Huang:2014xua}
\begin{equation}
2(m_{\chi} + \mu \sin 2\beta) \frac{1}{m_h^2} \simeq -\mu \tan\beta\frac{1}{m_H^2} \  .
\label{eq:BS}
\end{equation}
 Eq.~(\ref{eq:BS}) defines what we call a generalized blind spot in direct dark matter detection experiments. 
 It is clear from this expression that the  blind spot scenario demands $\mu < 0$. 
 Considering the case of heavy gluinos and scalar superpartners of the quarks and leptons, and assuming that the wino is significantly heavier than the bino (in practice we will assume  the relation implied by gaugino mass unification, $M_2 \simeq 2 M_1$), the generalized blind spot scenario can accommodate the right relic density in the well-tempered region~\cite{welltempered}, in which $M_1 \simeq |\mu|$, as well as the A-funnel region, in which $M_A \simeq 2 m_\no$ and the proper relic density~\cite{WMAP,planck} is obtained through resonant annihilation with the heavy Higgs bosons. 
 In this article
 we shall not analyze the case of wino-bino mixed dark matter that demands a high degree of degeneracy of the gaugino masses, $M_2 \simeq M_1$,
 and that leads to an extra suppression of the spin independent DDMD cross section due to the small Higgsino component of the dark matter candidates.
 For details of this case, see Ref.~\cite{Bramante:2015una}. In addition to Eq.~(\ref{eq:BS}), pure Higgsino or gaugino states, also lead to a large suppression to the DDMD~(see, for instance Ref.~\cite{Hill:2011be} and ~\cite{Bagnaschi:2016xfg}). In this work, we will concentrate on the region where $M_1 < M_2$, in which the thermal relic density can be naturally obtained for $m_\no$ of the order of weak scale.


\section{Direct Dark Matter Detection Constraints}
\label{sec:DDMDC}
\subsection{Allowed Parameter Space}

\begin{figure}[tbh]{
		\includegraphics[scale=0.5, clip]{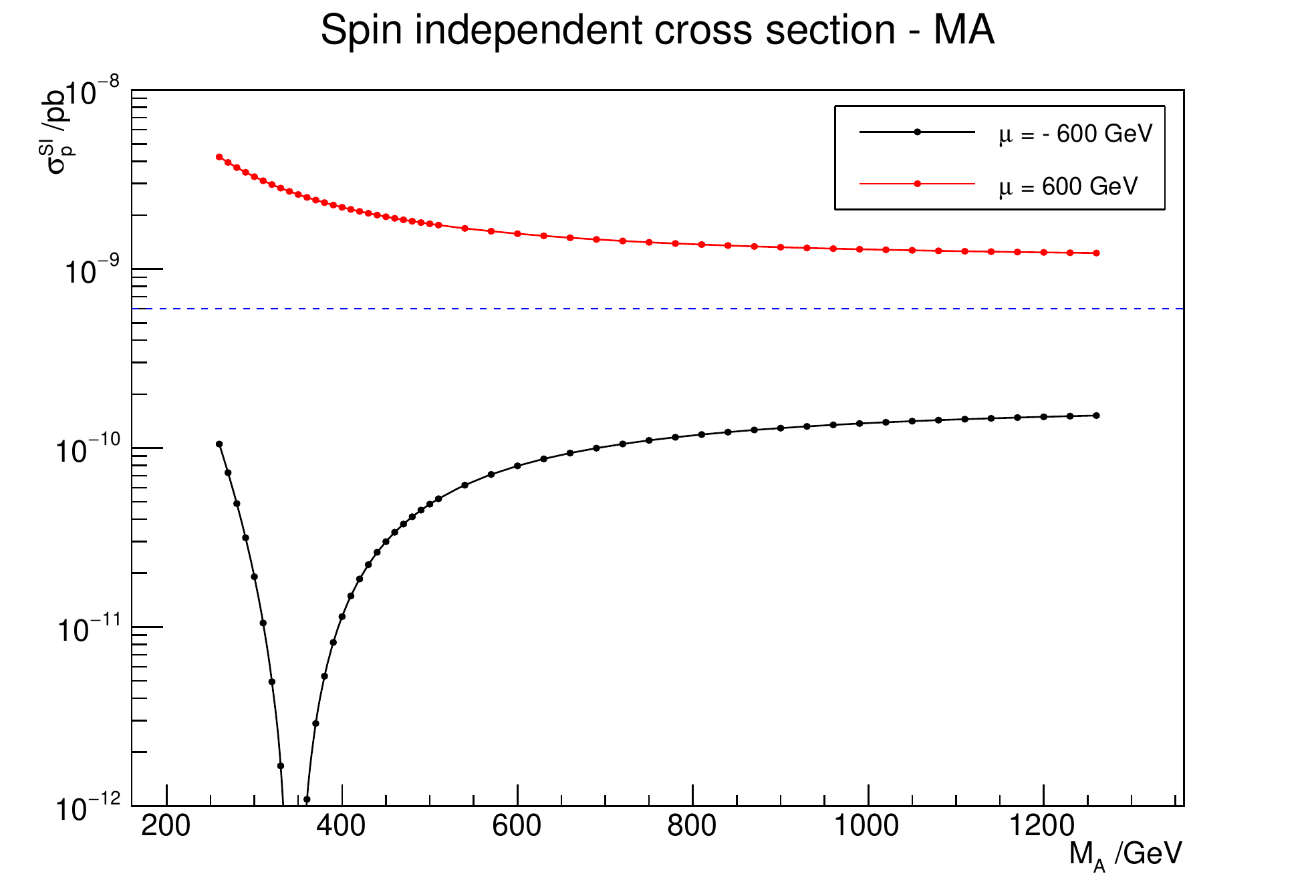}
		\caption{Spin independent scattering cross section for fixed $\tan\beta=7$, $|\mu|\ =600$ GeV, $M_1=400$ GeV. The black line is for $\mu<0$ and the red line is for $\mu>0$. The blue dashed line represents the LUX 2016 constraint for $m_\chi=400$ GeV. As $M_A$ increases the heavy Higgs becomes decoupled from the LSP, and $\sigma^{\text{SI}}_p$ approaches an asymptotic value. Note that the asymptotic value is significantly greater for $\mu>0$ than for $\mu<0$. When experimental limits drop below the asymptotic value of the $\mu<0$ branch, an upper bound and a lower bound on $M_A$ will be present, and we are forced closer to the blind spot.}
		\label{fig:sigma-MA}}
\end{figure}

\begin{figure}[tbh]{
	\includegraphics[scale=0.5, clip]{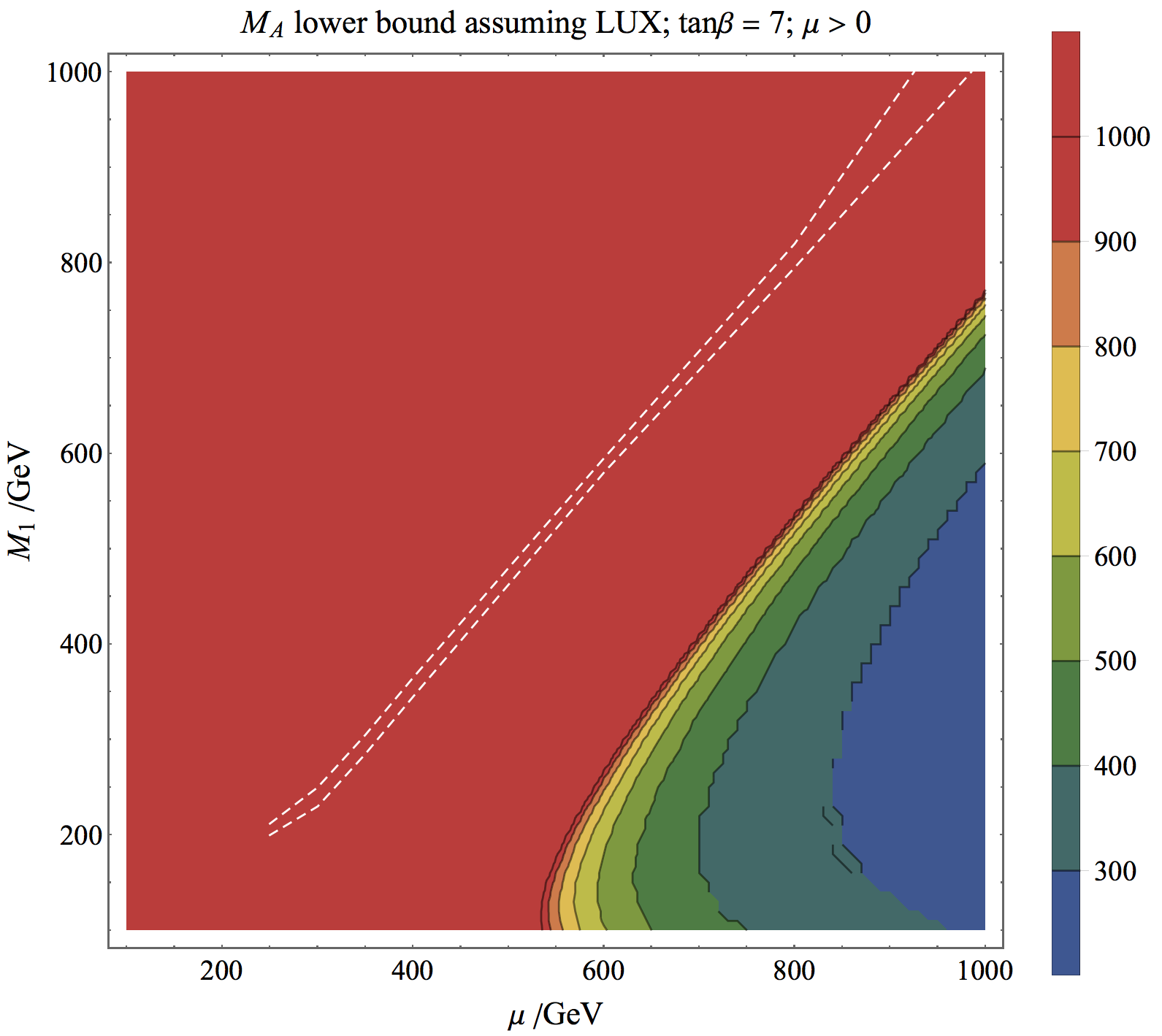}
	\caption{Lower bounds on $M_A$ due to 2016 LUX bounds for $\mu>0$, assuming the observed relic density in the whole parameter space. The value of $M_A$ is chosen to be at the minimum value allowed by the LUX bound, and is indicated by the color scale. In cases where the SI cross section is not allowed for all values of $M_A$, the lower bound is marked as infinity, corresponding to the red color. The region between the white dashed lines represents the well-tempered region, with the relic densities that differ from the observed value by less than 20\%. It can be seen that the well-tempered region is completely excluded. However, near the blue region away from the well-tempered region, the correct thermal relic density may still be achieved by resonant annihilation.}
	\label{fig:MAlbLUX_MUpos}}
\end{figure}

The expression of the SI cross section, Eq.~(\ref{eq:sig}), shows that, in general, the light Higgs and heavy Higgs contributions interfere destructively (constructively) for negative (positive) values of $\mu$. For negative $\mu$, scattering cross sections below the present LUX bounds are achievable even when the heavy Higgs boson is decoupled from the LSP as negative values of $\mu$ lead to a suppression in the neutralino Higgs coupling. For instance, for $\tan\beta=7$, $\mu=\pm600$ GeV, and $M_1=400$ GeV, Fig.~\ref{fig:sigma-MA} demonstrates that the scattering cross section may approach a relatively low asymptotic value in the limit $M_A\rightarrow\infty$ where the heavy Higgs is decoupled. The present LUX bound for $m_\chi\simeq M_1=400$~GeV is approximately $\sigma^{\text{SI}}_p<5\times10^{-10}$ pb \cite{luxfinal}, so for the example in Fig.~\ref{fig:sigma-MA} all cross section values obtained for  $\mu > 0$ are excluded. To quantify how much a particular point in the $\mu-M_1$ plane is excluded, we compute the minimal value of $M_A$ consistent with current LUX bound \cite{luxfinal} in Fig.~\ref{fig:MAlbLUX_MUpos}. In the red region, the lower bound on $M_A$ tends to infinity, indicating that the particular point is excluded for all $M_A$.  In applying the LUX bounds, we have implicitly assumed that the right relic density is obtained in all the parameter space, which could, for instance, demand a non-thermal contribution in large regions of parameter~\cite{Gelmini:2006pw}.
However, the exclusion region covers the entirety of the well-tempered region $M_1\simeq\mu$~\cite{Baer:2016ucr}, and one can only obtain the correct relic density via heavy Higgs mediated resonant annihilation near the blue region. We will comment on this case later, but since the SI cross section is in tension with the current LUX bound in the majority of region consistent with a observed thermal dark matter relic density, we shall focus our attention on the $\mu < 0$ case.

\begin{figure}[tbh]{
	\includegraphics[scale=0.4, clip]{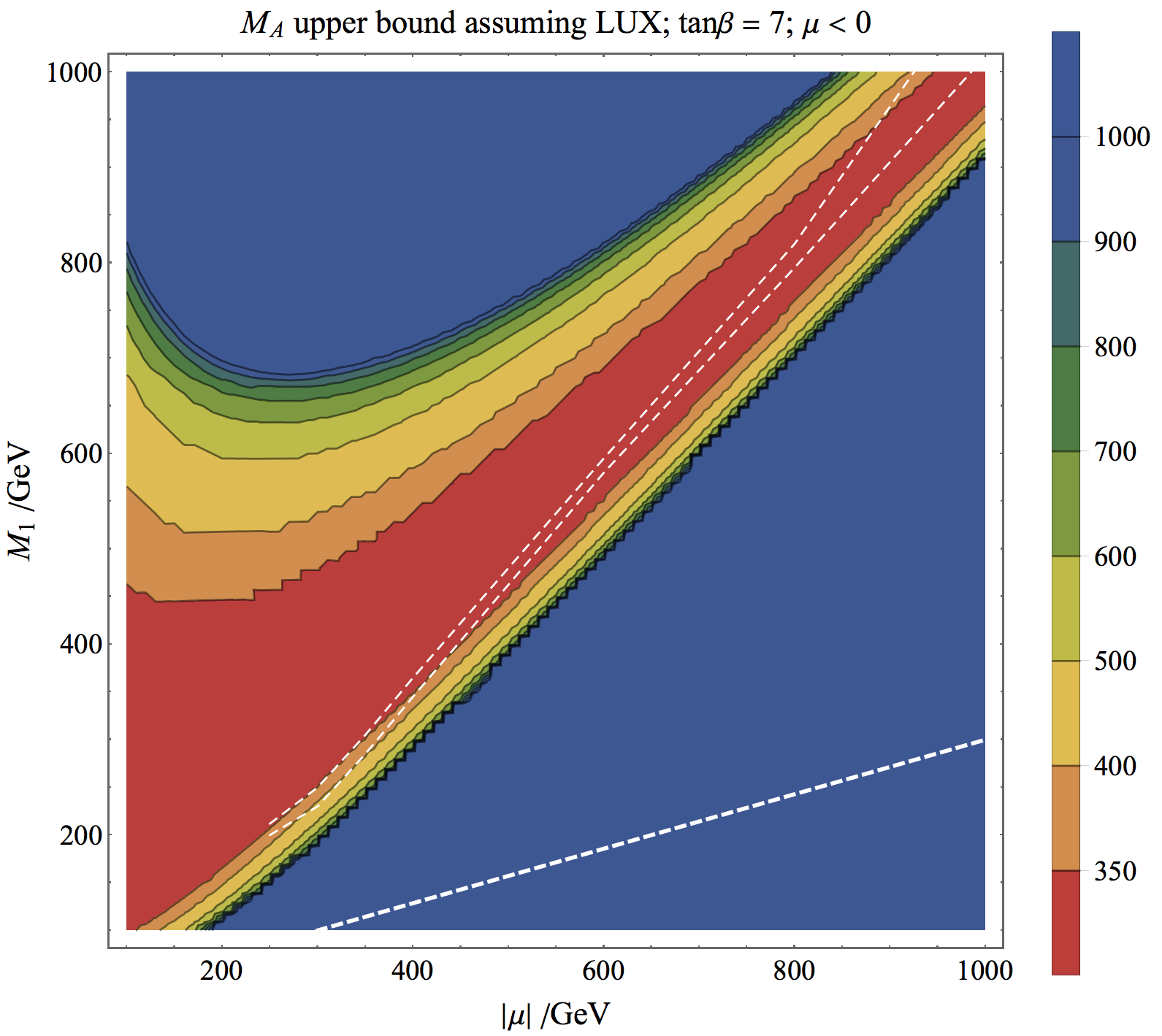}
	\includegraphics[scale=0.4, clip]{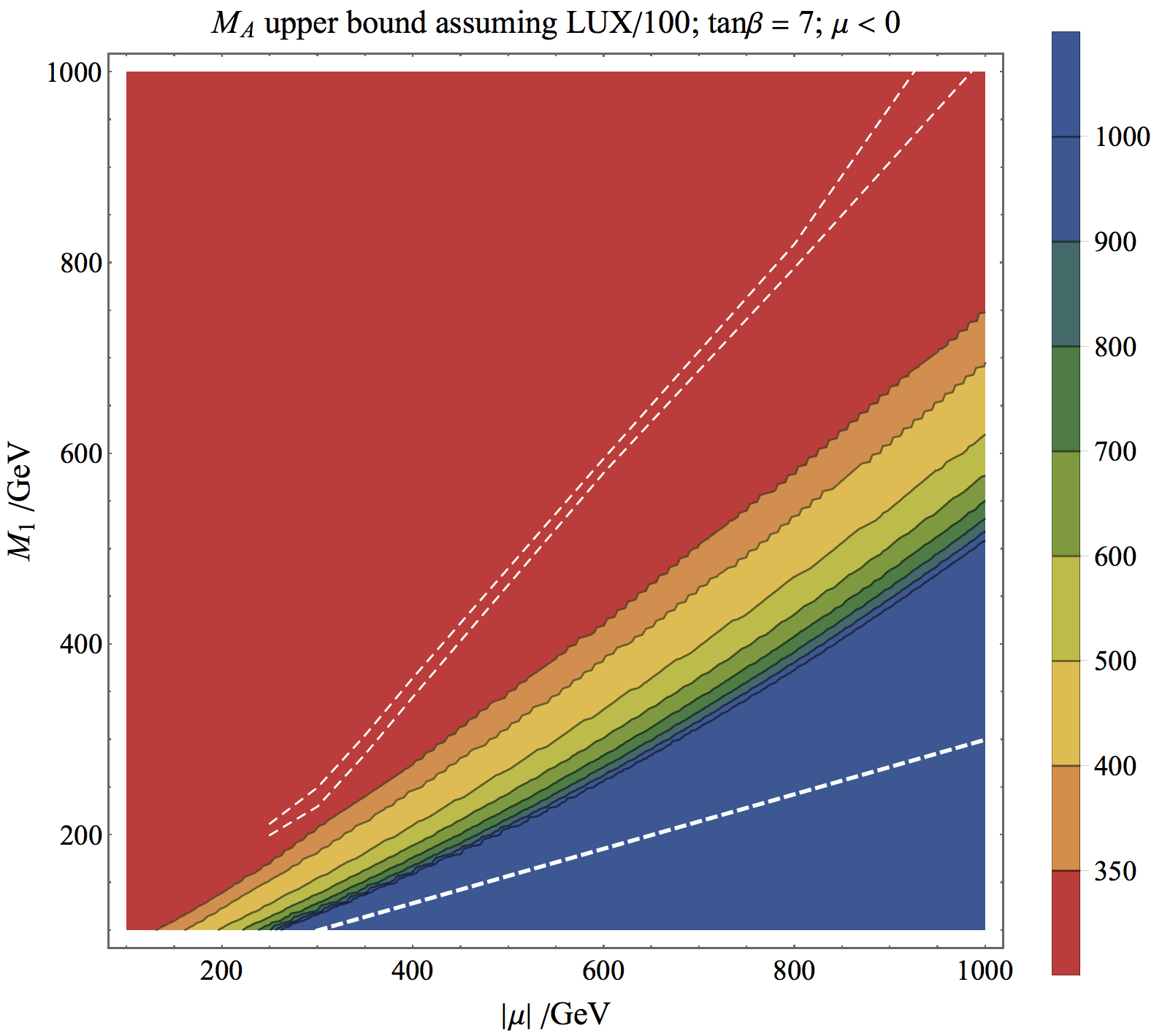}
	\caption{Upper bounds on $M_A$ due to 2016 LUX bounds and projected DDMD bounds 100 times stronger than LUX, respectively ($\mu<0$), assuming the observed relic density in the whole parameter space. The value of $M_A$ is chosen to be at the maximum value allowed by these bounds, and is indicated by the color scale. (Note that the color scheme differs from the previous plot such that the regions where the SI cross section is allowed as $M_A\rightarrow\infty$ is always shown in blue.) The region between the white dashed lines represents the well-tempered region. Under the strengthened bound a much larger portion of the $|\mu|-M_1$ plane is constrained. The dashed line below corresponds to where the left hand side of Eq.~\ref{eq:BS} is zero, corresponding to the vanishing of the neutralino coupling to the SM Higgs. Below this line the blind spot cannot be obtained since the left hand side of Eq.~\ref{eq:BS} becomes negative.}
	\label{fig:MAubLUX}}
\end{figure}

For the $\mu<0$ case, and assuming again the proper relic density in the whole parameter space, we compute the maximal value of $M_A$ consistent with current spin independent DDMD bounds in order to quantify the need for a destructive interference to reduce $\sigma_p^{\text{SI}}$ for the $\mu<0$ case. 
Fig.~\ref{fig:MAubLUX} shows that, contrary to the $\mu>0$ case, the present LUX bounds~\cite{luxfinal} only constrain the value of $M_A$ away from the decoupling limit close to the well-tempered region $M_1 \simeq |\mu|$, or for $|\mu| \ll M_1$. 

\begin{figure}[tbh]{
	\includegraphics[scale=0.5, clip]{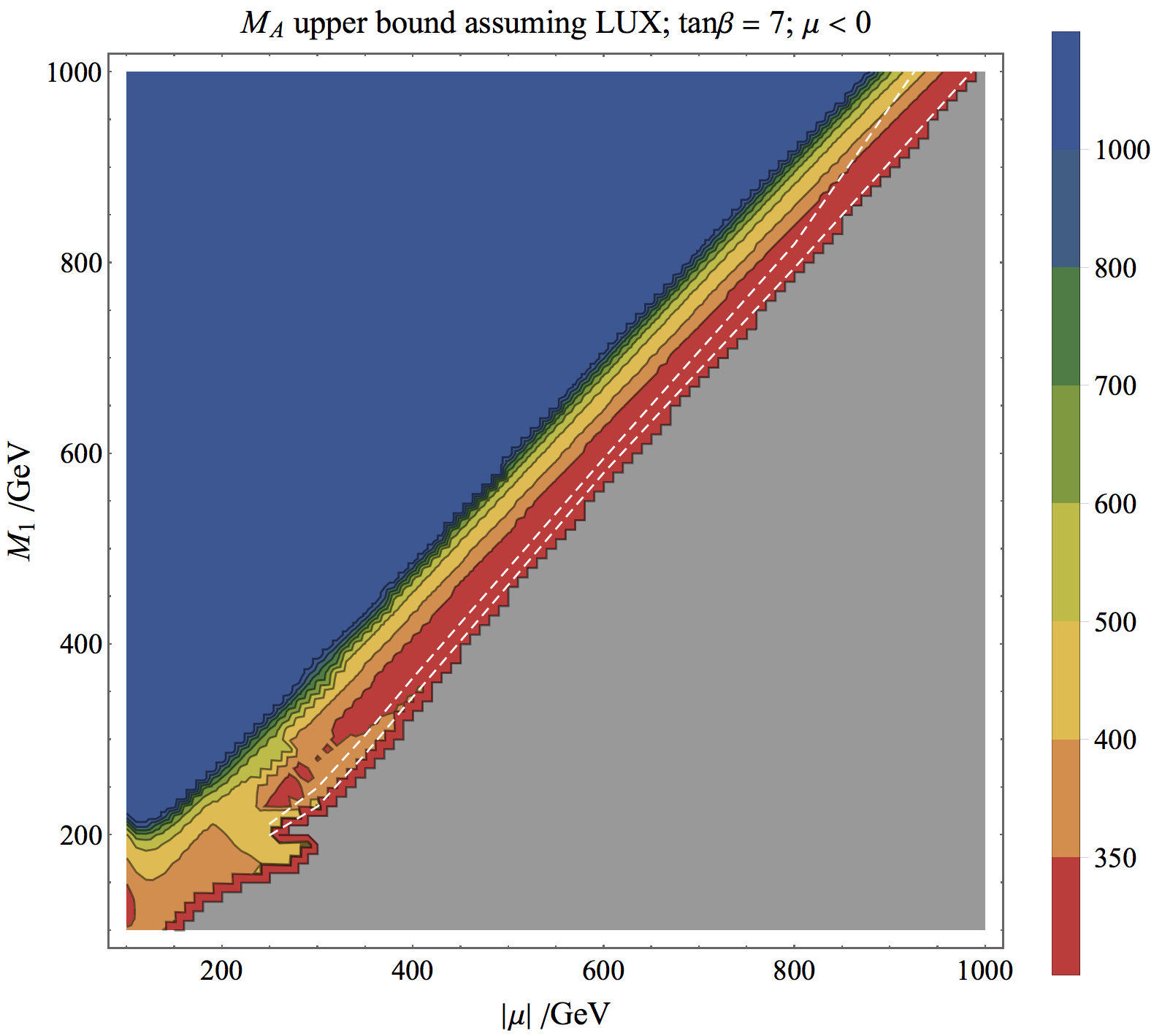}
	\caption{Upper bounds on $M_A$ due to 2016 LUX bounds, adjusted by the thermal WIMP relic density at each point in the plane. The strength of the LUX bound quickly decreases as one departs from the well-tempered region, since the WIMP relic density decrease quickly below the correct value. The gray region has relic density greater than 1.2 times the correct relic density, unless the neutralino mass is close to the resonant annihilation condition, 
	$m_A = 2 m_\no$, for which the proper relic density may be obtained and the upper bound becomes the one shown in Fig.~\ref{fig:MAubLUX}.}
	\label{fig:MAubLUXRdmod}}
\end{figure}

If one assumed a thermal origin of the relic density, then the above constraints would be modified. In the upper left region of the plot, the thermal neutralino dark matter is under-abundant (see Fig.~\ref{fig:RD-muM1} for the thermal relic density), so we assume there is another component, for instance, the axions, that contributes to the relic density. In this case, the LUX bound should be rescaled according to the thermal relic density of $\no$, thereby relaxing the upper bound on $M_A$, as shown in Fig.~\ref{fig:MAubLUXRdmod}. The proper relic density may be obtained thermally in the well-tempered region (inside the white dashed band in Fig.~\ref{fig:MAubLUX}), for which LUX imposes an upper bound on $M_A$. This upper bound is of order of 350~GeV or smaller for $\tan\beta = 7$ and large values of the neutralino mass. However, it raises to values of order 400~GeV for neutralino masses of the order of 300~GeV. 

The proper thermal relic density may be also obtained to the right of the well-tempered region, in the so-called A-funnel region, by setting $M_A \simeq 2 m_\no$ such that the heavy Higgses mediate the resonant annihilation of the LSP, reducing the relic density to the correct value. Thus, points in the parameter space are allowed by LUX and relic density consideration when the upper bound on $M_A$ is larger than or on the order of $2 m_\no$, corresponding to the blue and green regions in Fig.~\ref{fig:MAubLUX} with $M_1 < |\mu|$. For sufficiently small values of $M_1$ the value of the amplitude due to the exchange of the heavy CP-even Higgs may be sufficiently large to induce an increase of the cross section toward values restricted by LUX, as seen on the left of Fig.~\ref{fig:sigma-MA}. As we will show below, this situation only occurs for very small values of $M_1$. Larger values of $M_1$, of the order of the weak scale, would only be restricted if future SI DDMD experiments fail to see a signal.

It is interesting to investigate the region to be probed by future DDMD experiments. In case of no detection, future experiments will push the experimental limits below the decoupled scattering cross section in greater regions of the $\mu-M_1$ plane. In particular, the projected bounds of the LZ experiment are approximately 100 times stronger than those from the LUX experiment~\cite{lz}. Fig.~\ref{fig:MAubLUX} reveals that, assuming a dark matter density consistent with the observed one,  these stronger bounds would constrain $M_A$ in the entire region left of the well-tempered region, and in part of the region to the right as well.  As before, if a thermal origin of the dark matter
relic density is assumed, the well-tempered region may be achieved, but the upper bound on $M_A$ would become smaller than about 300~GeV.

\begin{figure}[tbh]{
	\includegraphics[scale=0.6, clip]{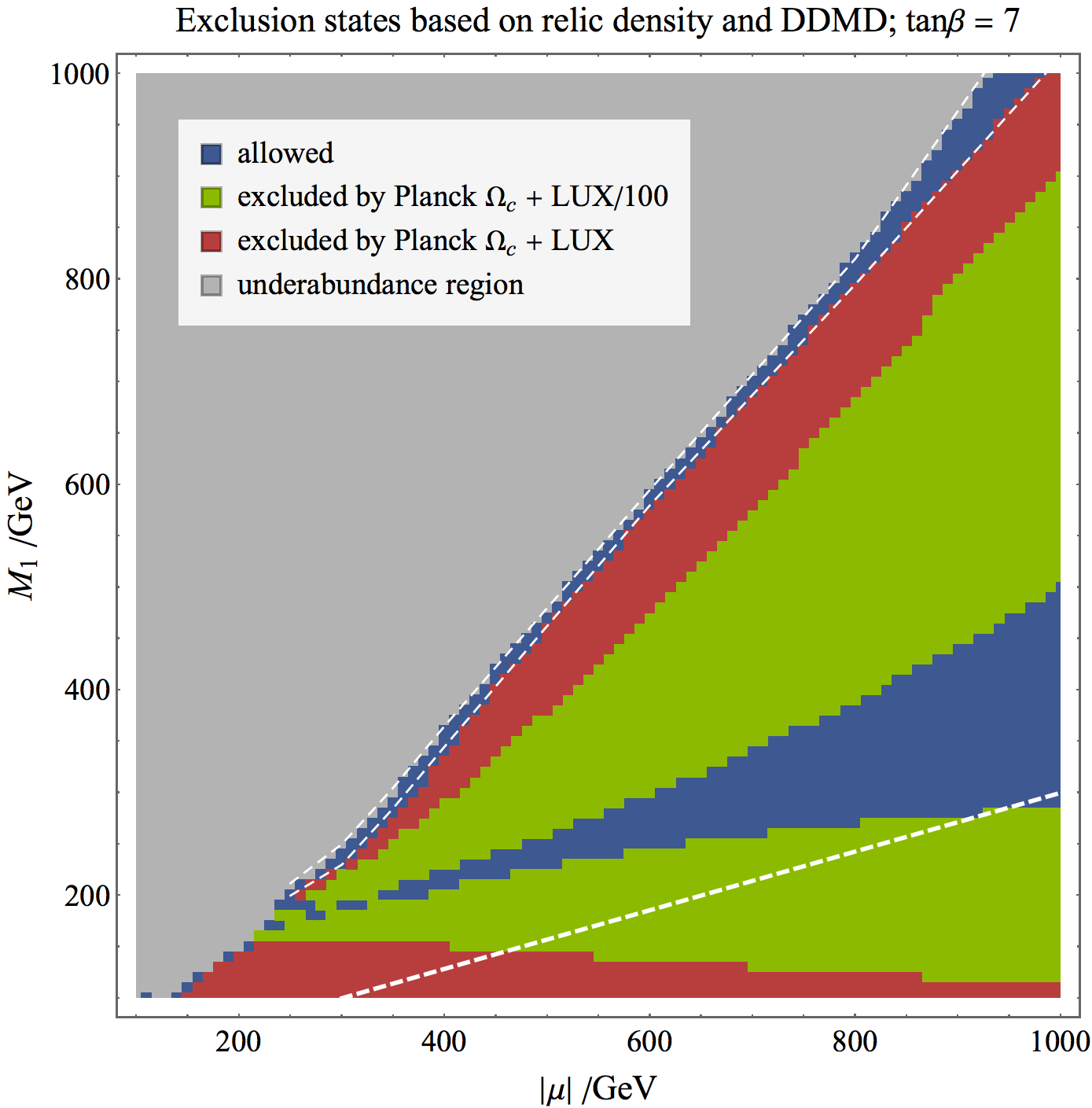}
	\caption{Constraints on the $|\mu|-M_1$ plane under relic density constraints and the present and projected DDMD constraints. The well-tempered region ($\mu \simeq M_1$) naturally attains the correct relic density, while the region below may attain the correct value if $M_A$ is tuned to mediate resonant annihilation. The required value of $M_A$ is constrained by the LUX and 100 times strengthened LUX bound on SI cross section. The blue region is allowed by the 100 times strengthened LUX bound; the blue and the green regions are allowed under current LUX bound. Note that the boundaries of the LUX constraint (red) and of the LUX/100 constraint (green) above the blue region correspond to the boundaries in Fig. \ref{fig:MAubLUX} where the upper bound on $M_A$ is quickly lifted to infinity. The constraints below the blue region are due to the overcompensation in the scattering cross section from the heavy Higgs contribution.}
	\label{fig:Rd_constraints}}
\end{figure}

A more complete description of the exclusion state of the A-funnel region takes into account the upper bound on $M_A$ presented above as well as the lower bound due to the overcompensation of the heavy CP-even Higgs contribution. As mentioned before, the region allowed by LUX and relic density considerations roughly correspond to the blue and dark green region in Fig.~\ref{fig:MAubLUX}, where the required value for resonant annihilation $M_A\simeq2M_1$ is below the upper bound set by LUX. (For $\mu>0$, the correct relic density can be achieved near the blue region of Fig.~\ref{fig:MAlbLUX_MUpos}, where the required value is above the lower bound set by LUX.) The constraints from both sides are summarized in Fig.~\ref{fig:Rd_constraints}, which shows the exclusion states of the $|\mu|-M_1$ plane under present and projected DDMD constraints together with the relic density consideration.
Below the viable well-tempered region, the exclusion states are determined by fixing $M_A$ close to the resonant value $2 m_\no$ and comparing the SI cross section with the current and future bounds.
The resulting bounds in this region combine the previous constraint on $M_A$ away from the decoupling limit (upper bound) with the constraints from below. It can be seen that the present LUX bound leave the parameter space relatively open, while the projected (100 times strengthened) bound would considerably constrain the region in which resonant annihilation can be employed to obtain the correct relic density. In regions where the WIMP relic density is under-abundant, the upper bounds on $M_A$ lifts up quickly if one adjust the LUX bound to match the WIMP relic density specific to each point in the $\mu-M_1$ plane, as shown in Fig.~\ref{fig:MAubLUXRdmod}, opening up space for studies on mixed dark matter origin. We shall not concentrate on this scenario.


\subsection{LZ Reach and Blind Spots}
\label{sec:num}

The lack of observation of a signal at the LZ experiment would constrain us to a narrow region of allowed parameter space for thermal 
dark matter, namely the A-funnel region displayed in Fig.~\ref{fig:Rd_constraints}, plus the well-tempered region for values of $M_A$ consistent with the
upper bound obtained in Fig.~\ref{fig:MAubLUX}. The reach of LZ goes far beyond the natural values of the spin independent cross 
section for values of the gaugino and Higgsino masses of order of the weak scale, and therefore pushes the parameters towards the
blind spot values.  Alternatively, one could consider the event of an LZ detection of Dark Matter in the currently allowed range. 
In order to fix ideas and show the complementarity of different search methods in detecting dark matter, we shall assume a detection of dark
matter with a spin independent cross section of the order of $\sigma^{\text{SI}}_p = 10^{-11}$~pb,
which is about 100 times lower than the LUX bound for $m_\chi = 500$ GeV, and  within the sensitivity of the LZ  experiments for WIMP masses 20~GeV$\simlt m_\chi \simlt$~500~GeV. Thus, for this range of WIMP masses, of order of the weak scale, future DDMD experiments will either detect dark matter or determine the proximity to the blind spot scenario, and probe it in such a case. Since, in addition, the blind spot scenario requires a specific correlation between the Higgs and neutralino masses, it can also be efficiently probed by spin dependent and indirect dark matter detection experiments, as well as by searches for Higgs and electroweakinos at the LHC. 

The parameter space in which these small spin independent cross sections are achieved is well captured by the phenomenological MSSM (pMSSM) parameter space~\cite{pmssm}, but we have reduced its dimensionality to conduct a feasible study and to concentrate on critical variables. For instance, sfermion masses are held constant at 2~TeV, above experimental constraints, as they have little impact on the determination of the neutralino relic density and on DDMD experiments unless their masses are comparable to that of the LSP.  By doing this, we are eliminating the interesting possibility of dark matter co-annihilation with scalar superpartners (see, for example, Refs.~\cite{Ellis:1999mm,Balazs:2004bu,Ellis:2001nx, Ajaib:2011hs,Ibarra:2015nca}), which is of phenomenological interest. We reserve the study of this case for future work.
 
In our analysis the trilinear coupling constants are fixed to be zero except for $A_t$ , which is taken at a value of order $2~M_S = 2\sqrt{M_{\tilde t_1}M_{\tilde t_2}}$ to obtain the proper 125 GeV Higgs mass~\cite{Draper:2013oza,Vega:2015fna,Lee:2015uza}. The values of $M_2$ and $M_3$ are held well above $M_1$ so that the heavier electroweakinos do not interfere with the annihilation of the LSP.  As mentioned before,  we fixed $M_3 = 2$~TeV and chose $M_2$ by imposing the gaugino mass unification $M_2=2 \ M_1$ for simplicity, but our results are general whenever $M_2 \gtrsim 2 \ M_1$.  As stressed before,
the constraints on the parameter space become stronger for smaller values of $M_2$ and therefore we shall concentrate on the case of a Bino-like
neutralino that leads to a larger allowed parameter space and also allows the obtention of a thermal relic density in larger regions of parameter space.

The four remaining parameters\textemdash $\tanb$, $M_1$, $\mu$, and $M_A$ \textemdash are critical to our model. For each combination of $\tanb$, $M_1$ and $\mu$, we select $M_A$ to obtain a  cross section $\sigma_p^{SI}$ smaller than the required bound on this quantity. In practice, as
an example, we shall allow $M_A$ to vary in the range consistent with $\sigma_p^{SI} \leq 10^{-11}$ pb. Hence, the boundaries of this region will be
consistent with a potential measurement by LZ, while the central point would be close to the parameters leading to the blind spot scenario.  
We shall focus our study on the region $5\leq\tanb\leq15$, that may accommodate the proper Higgs boson mass within the MSSM, and where our parameter space is left relatively open by the LHC $H\rightarrow\tau\tau$ and electroweakino (EWino) searches, and other collider constraints.

\begin{figure}[tbh]{
	\includegraphics[width = \textwidth, clip]{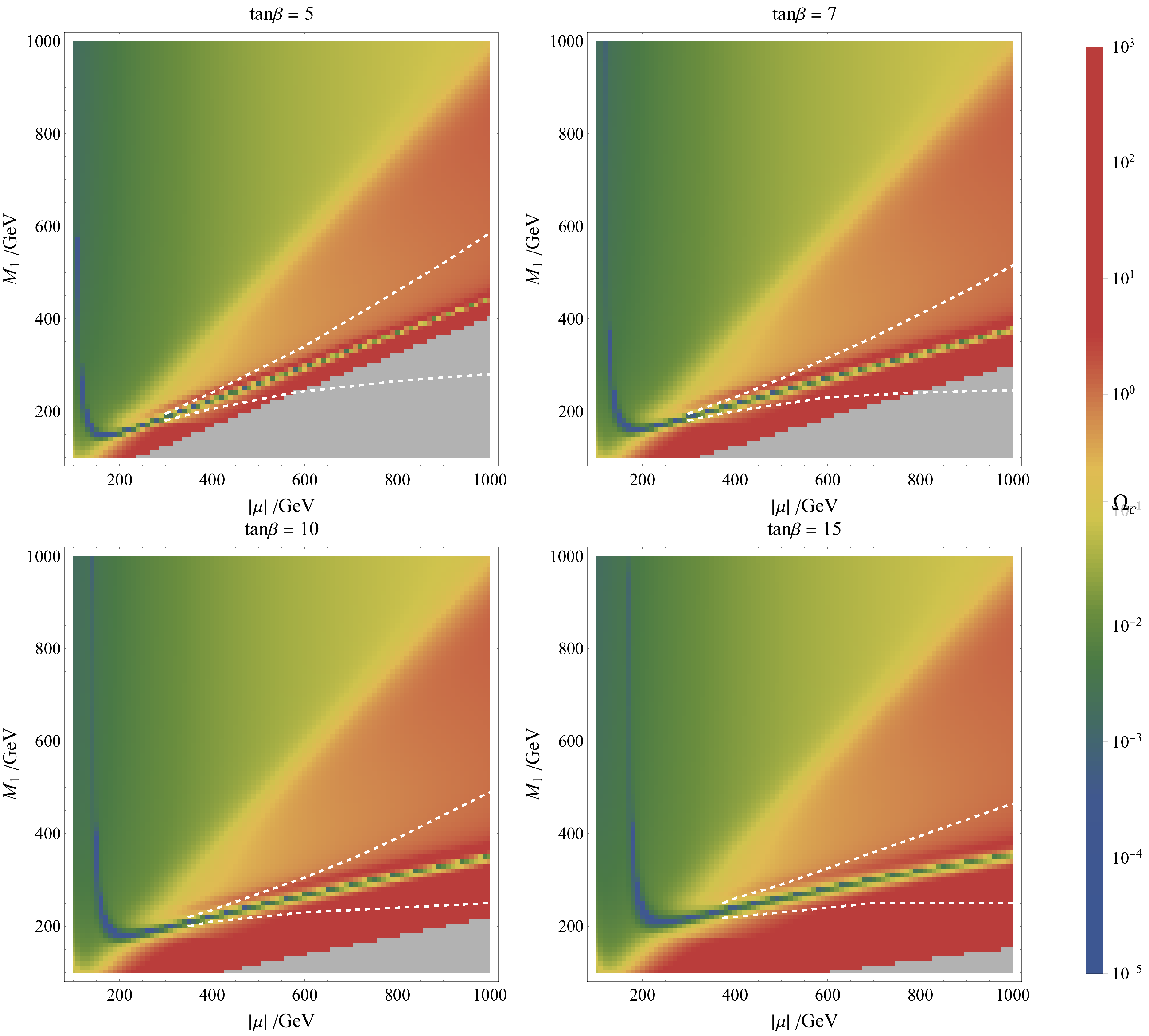}
	\caption{Thermal relic density shown in color on $|\mu|-M_1$ plane for various $\tanb$. $M_A$ is taken to be at the center of blind spot (maximum cancellation). Note that $\mu$ is always negative for the blind spot to occur. The yellow region is consistent with the observed relic density. In the regions between the white dashed lines, $M_A$ can be adjusted to mediate resonant annihilation while keeping $\sigma_p^{SI}<10^{-11}$~pb. Blind spots are not achieved in the gray area since the left hand side of Eq.~\ref{eq:BS} becomes negative and destructive interference cannot happen. In this region, the $\sigma_p^{SI}<10^{-11}$~pb requirement does not set an upper bound for $M_A$ but only a lower bound, though it is still possible to tune $M_A$ to achieve resonant annihilation for $m_\chi$ high enough.}
	\label{fig:RD-muM1}}
\end{figure}

We have used MicroOMEGAs (with SuSpect 2.41)  to calculate the spectrum, SI and SD DDMD cross sections and corresponding relic densities~\cite{micromegas}. The thermal relic density is displayed on the $\mu-M_1$ plane in Fig.~\ref{fig:RD-muM1} for various values of $\tanb$, with $M_A$ fixed at values 
consistent with the blind spot center (maximal destructive interference). The yellow color indicates that the region has the relic density consistent with the observed one~\cite{WMAP,planck}. It can be clearly seen that the desired region consists of two branches: the well-tempered region $|\mu| \simeq M_1$ in the upper branch and the A-funnel region $|\mu| \simeq 2M_1$ in the lower branch. The resonant annihilation with the heavy Higgs is in fact so strong that the relic density rapidly decreases towards the center of the A-funnel region, so the correct relic density is only achieved on the two sides. However, since $M_A$ is allowed to vary in a small range such that $\sigma_p^{SI}<10^{-11}$~pb, the correct relic density can be attained in a wider region (between the white dashed lines in Fig.~\ref{fig:RD-muM1}) by fine-tuning $M_A$. 
It can also be seen from Fig.~\ref{fig:RD-muM1} that both the well-tempered region and the A-funnel region in the blind spot scenario continues on almost  linearly to $|\mu| \simeq 1$ TeV. These branches are approximated as piecewise linear functions when casting the collider constraints on the $\tanb-M_A$ plane in section IV.

\section{LHC Constraints}
\label{sec:LHC}

In this section we concentrate on the region of parameters consistent with $\sigma_p^{SI} \leq 10^{-11}$ pb.
Recent LHC 13 TeV data reveals no signal of any BSM particles. The new exclusion limits from the ATLAS and CMS collaborations are used to constrain the parameter space of the blind spot scenario. In our region of interest where $\tanb$ is between 5 and 15, the $H, A\rightarrow \tau\tau$ searches~\cite{CMS:htautau2016,ATLAS:Htautau} offer the most stringent constraints. Electroweakino searches at CMS~\cite{CMS:EWino2016} provide additional constraints in the region of small $M_1$ and $|\mu|$, where $\widetilde \chi_1^0$ is especially light. In subsection A we examine constraints from the CMS and ATLAS $H,A \rightarrow \tau \tau$ searches, followed by an analysis of constraints from the $\widetilde \chi_2^0 \widetilde \chi_1^\pm \rightarrow WZ \widetilde \chi_1^0 \widetilde \chi_1^0$ channel in subsection B. Overall, we find that the well-tempered region is completely excluded for $\tan \beta \geq 7$, and the A-funnel region is only partially excluded for the larger values of $\tan \beta$. The region of small $m_\no$ tends to be in tension with electroweakino searches. 

\subsection{$H \rightarrow \tau \tau$ Search}
\label{subsec:Hsearch}

We consider production of the heavy Higgses $H$ and $A$ (either of which is denoted by $\phi$) by means of gluon-gluon fusion ($gg\phi$) and b-associated production ($bb\phi$), followed by a decay into two $\tau$ leptons. Recent reports from CMS and ATLAS~\cite{CMS:htautau2016,ATLAS:Htautau} provide 2-dimensional 95\% confidence level (CL) upper limits on parameters related to these decays. CMS puts bounds on $\sigma(gg\phi) \times BR(\phi \rightarrow \tau \tau)$ with respect to $m_\phi$ and $\sigma(bb\phi) \times BR(\phi \rightarrow \tau \tau)$, while ATLAS puts bounds on $\sigma_{\text{tot}} \times BR(\phi \rightarrow \tau \tau)$ with respect to $m_\phi$ and $f_b = \sigma(bb\phi)/\sigma_{\text{tot}}$. Their bounds are given for discrete $M_A$ (and discrete $f_b$ for ATLAS), so we linearly interpolate to find bounds at arbitrary  values. While the CMS and ATLAS bounds consider the production and decay of either $H$ or $A$, these processes are experimentally indistinguishable since $M_A \simeq M_H$ in our model, so we sum the cross section times branching ratio and compare these summed values to the experimental limits. We use FeynHiggs 2.12.0~\cite{FeynHiggs1,FeynHiggs2,FeynHiggs6} to compute the relevant cross sections and branching ratios for points in the blind spot scenario. 

\begin{figure}[tbh]{
	\includegraphics[width = 3.68cm]{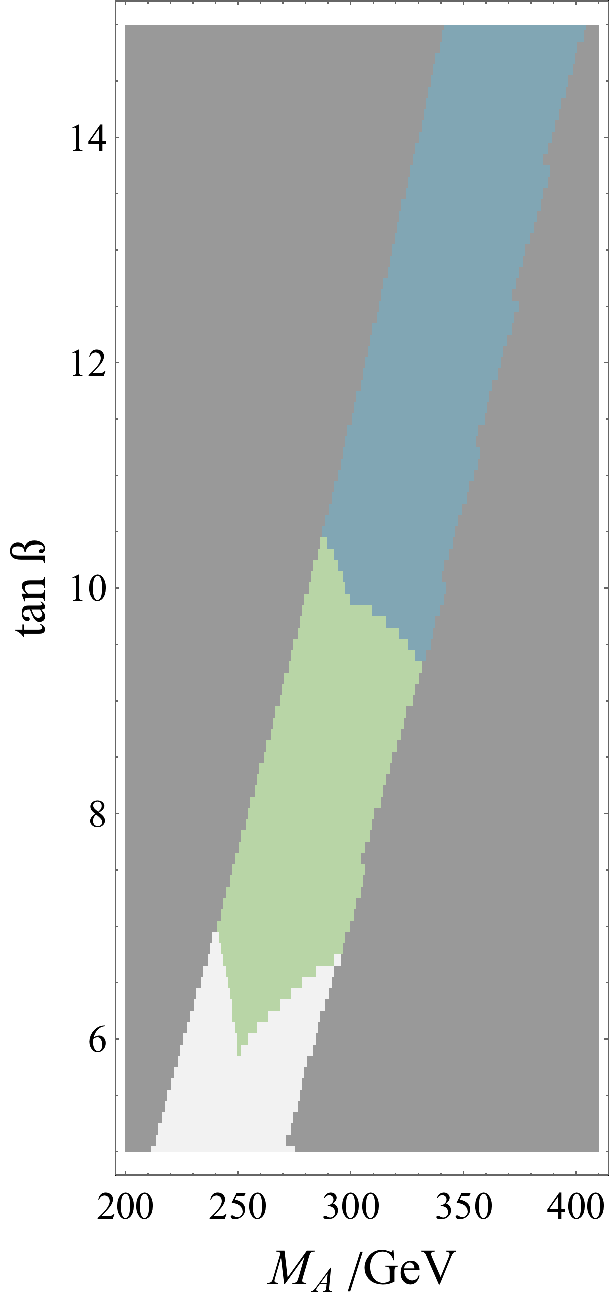}
	\includegraphics[width = 12cm]{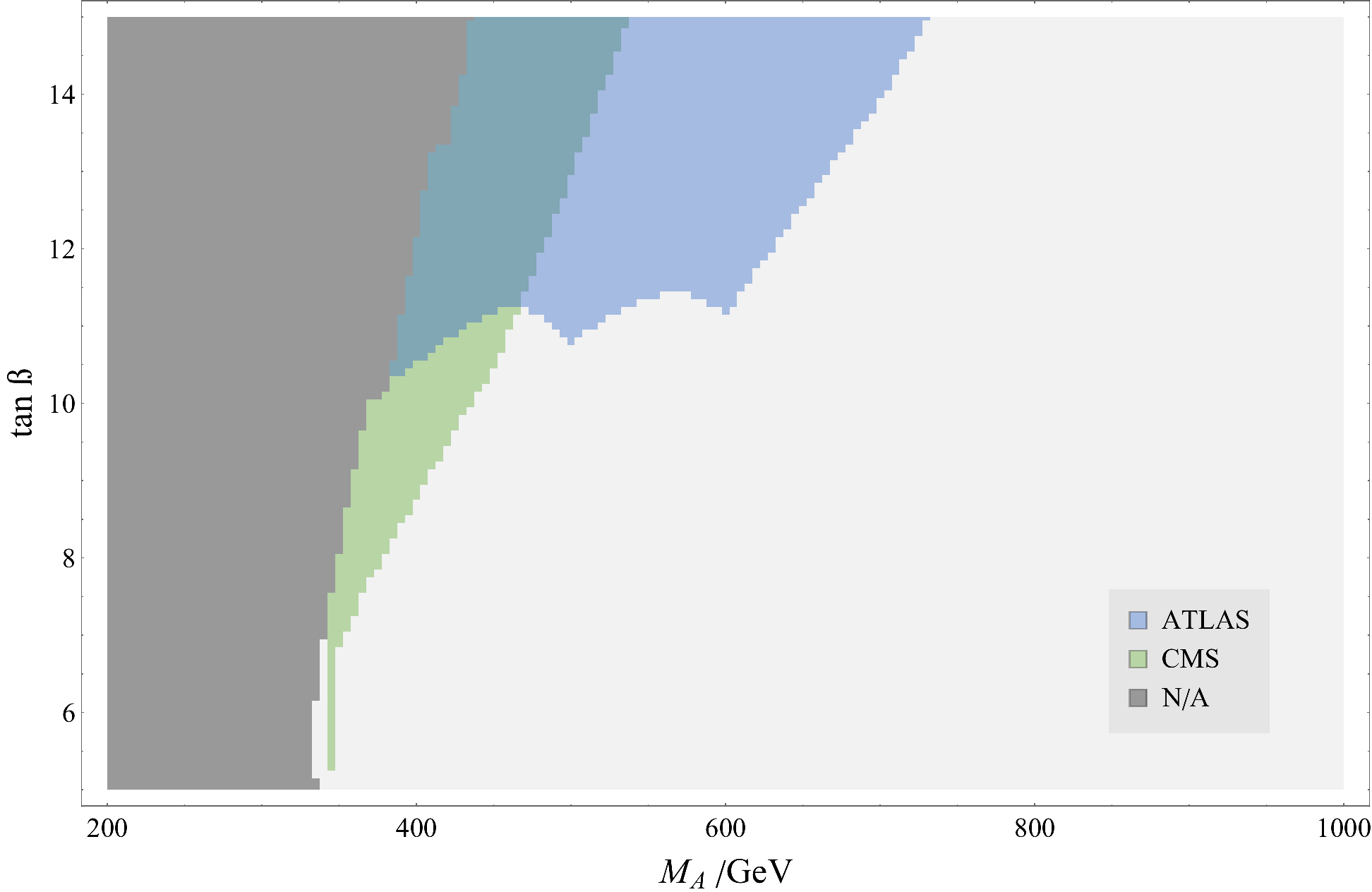}
	\caption{Exclusion bounds on the well-tempered region and the A-funnel region in the $\tanb-M_A$ plane. The well-tempered region represented in the left panel and the A-funnel region in the right panel. The colored regions are excluded at $95\%$ CL by the ATLAS or CMS results. The dark gray regions are not consistent with $\sigma_p^{SI} \leq 10^{-11}$~pb.}
	\label{fig:tanbMA}}
\end{figure}

Piecewise linear approximations are made for $M_1$ as a function of $|\mu|$ in the well-tempered and A-funnel branches for $M_1, |\mu|\; > 200$~GeV, based on Fig. \!\ref{fig:RD-muM1}. We shall consider only $M_1, |\mu|\; \gtrsim 200$~GeV in this subsection, leaving an analysis of the $M_1, |\mu|\; \lesssim 200$~GeV case for subsections $B$ and $D$.
Cross sections and branching ratios are computed at points along these approximations with $M_A$ chosen to be at the center of the blind spot, and are checked against the bounds in the ATLAS and CMS reports described above. The excluded regions in the $\tan\beta-M_A$ plane 
are shown in Fig.~\!\ref{fig:tanbMA}. 

\begin{figure}[tbh]{
	\includegraphics[width = 8.16cm, clip]{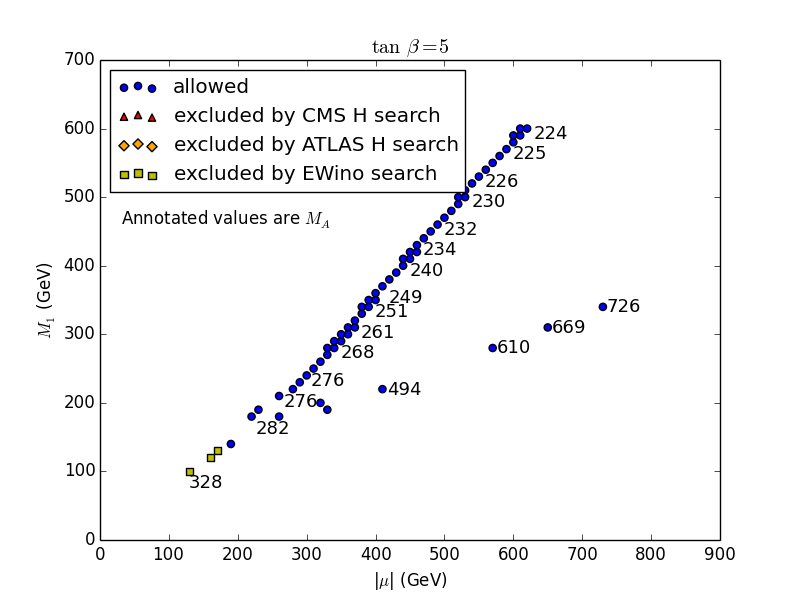}
	\includegraphics[width = 8.16cm, clip]{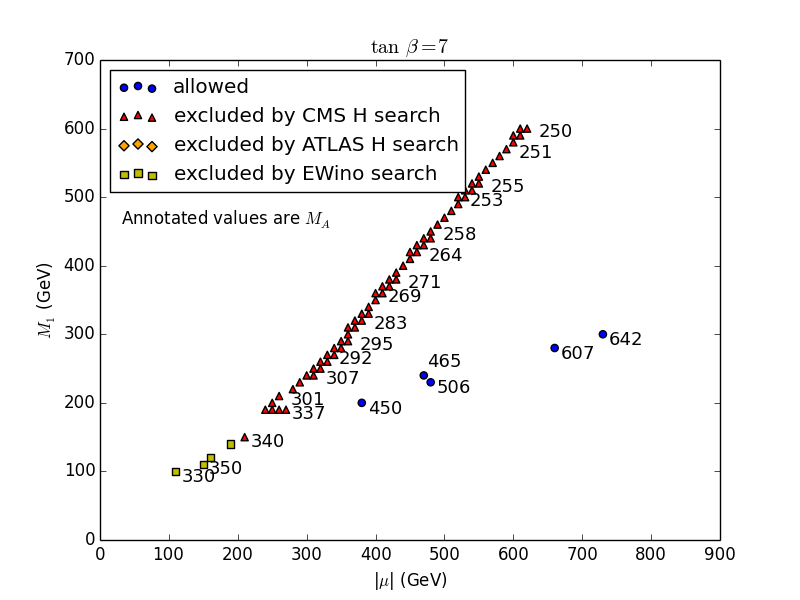}
	\includegraphics[width = 8.16cm,clip]{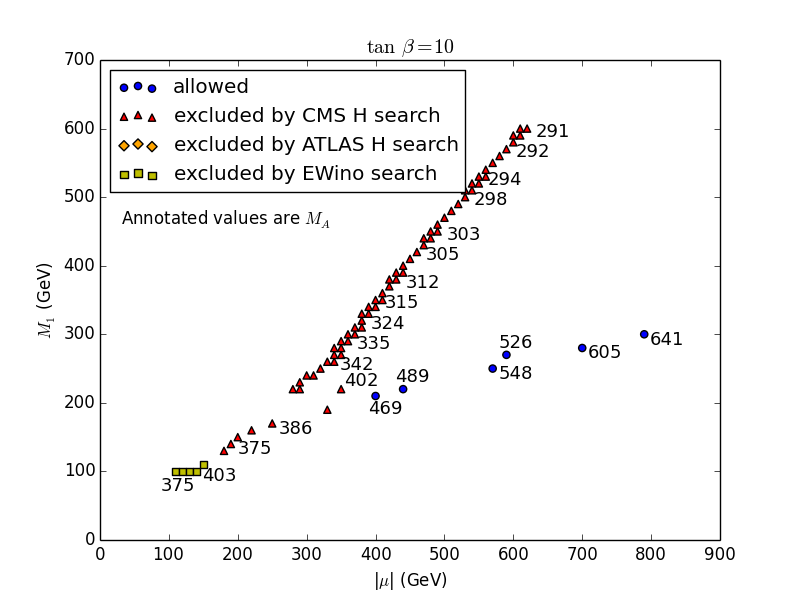}
	\includegraphics[width = 8.16cm,clip]{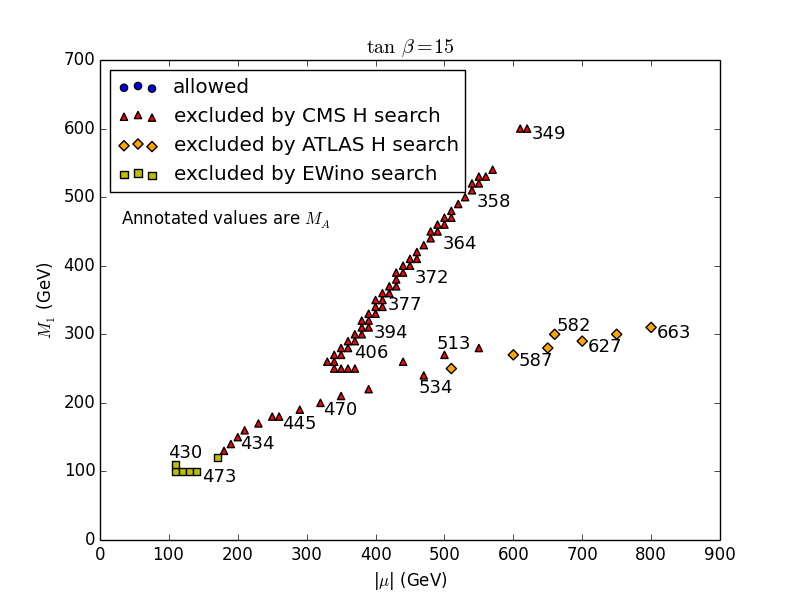}
	\caption{Net exclusion status of the well-tempered region and the A-funnel region. Each data point represents a point with the proper relic density. Data points are first checked for exclusion by CMS, then by ATLAS, then by the CMS electroweakino searches (see the following section), with the color-coding of each data point corresponding to the method by which it is first excluded. Values of $M_A$ are labeled next to selected data points. The sparseness of points in the A-funnel region reflects its narrowness relative to the well-tempered region, as seen in Fig. \!\ref{fig:RD-muM1}.}
	\label{fig:MA-muM1}}
\end{figure}

We see that at $\tan \beta = 6$ the well-tempered region is partially  excluded, and for $\tan \beta \geq 7$ it is completely ruled out. Parts of the A-funnel region remain available at all $\tan \beta$, though the ATLAS results restrict large portions of parameter space for $\tan \beta > 10$. To reconcile Fig. \!\ref{fig:RD-muM1} and Fig. \!\ref{fig:tanbMA}, we present the exclusions in the $|\mu|-M_1$ plane for blind spots with the proper relic density in Fig. \!\ref{fig:MA-muM1}. The data set used in Fig. \!\ref{fig:MA-muM1} is the same as for the yellow region in  Fig. \!\ref{fig:RD-muM1}, so each data point has $M_A$ chosen to be at the center of the blind spot,  considering data points with the correct dark matter density. Again, we see that the well-tempered region is excluded for $\tan \beta \geq 7$, and the A-funnel region begins to be excluded as well as $\tan \beta$ increases. All data points shown in the A-funnel region are excluded by ATLAS for the $\tan \beta = 15$ plot, but the figure may be extended to reach allowed regions of parameters at higher $|\mu|$ and $M_A$. Fig. \!\ref{fig:MA-muM1} also shows that $M_A$ tends to be smaller in the well-tempered region, resulting in a higher production cross section, hence the greater degree of exclusion. On the other hand, as $M_A$ increases in the A-funnel region, the $\phi \rightarrow \tau \tau$ branching ratio decreases as additional decay channels (notably the $\phi \rightarrow t\overline t$ channel) are opened and enhanced, resulting in a weaker exclusion limit.

\begin{figure}[tbh]{
	\includegraphics[width = 8.16cm, clip]{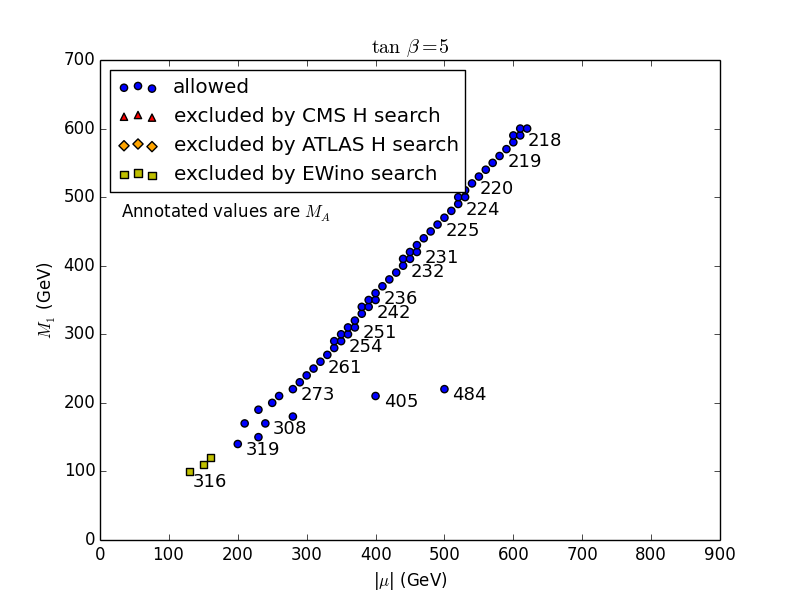}
	\includegraphics[width = 8.16cm, clip]{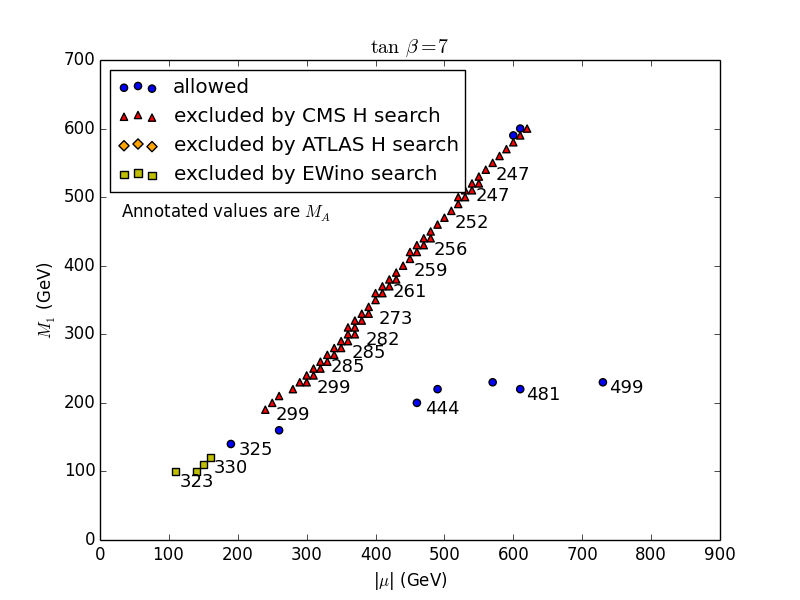}
	\includegraphics[width = 8.16cm,clip]{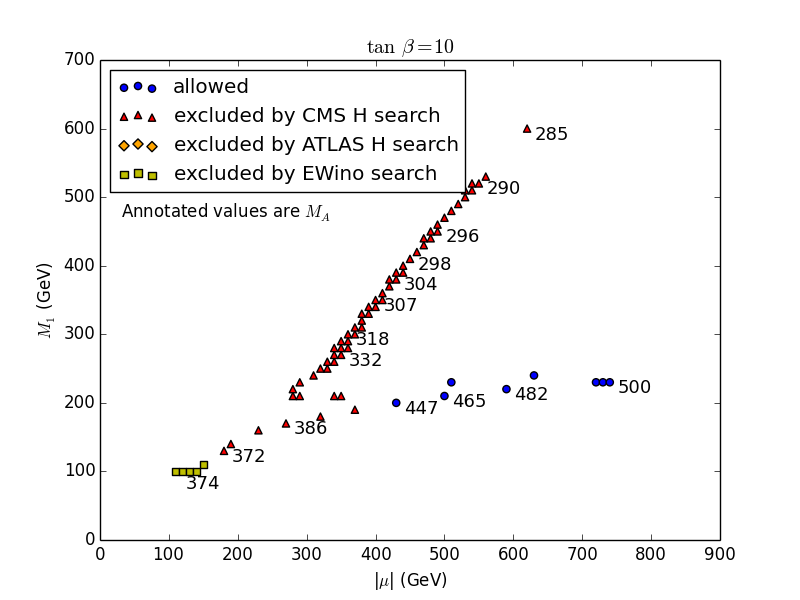}
	\includegraphics[width = 8.16cm,clip]{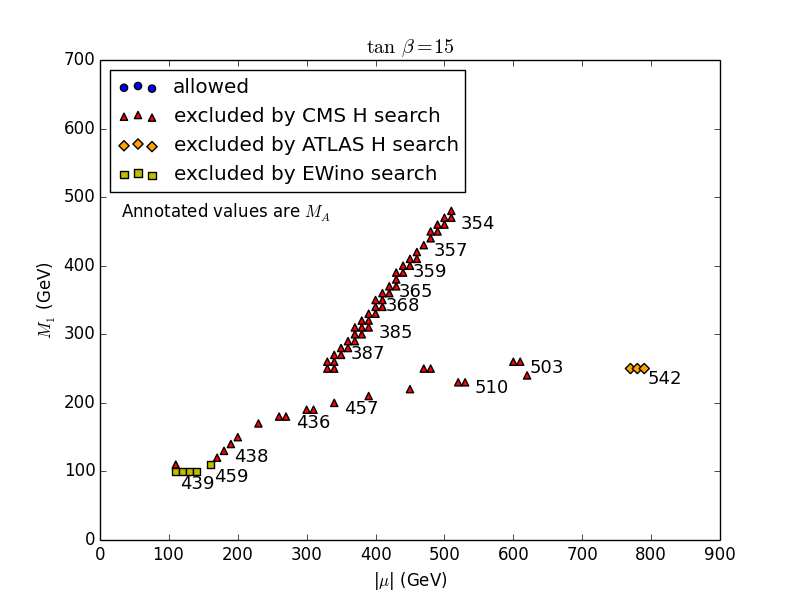}
	\caption{Plot analogous to Fig. \!\ref{fig:MA-muM1} with $M_A$ chosen at the lower boundary  consistent with $\sigma_p^{SI} = 10^{-11}$~pb . Only exclusions from the CMS $\phi \rightarrow \tau \tau$ search are shown.}
	\label{fig:MA-muM1L}}
\end{figure}

\begin{figure}[tbh]{
	\includegraphics[width = 8.16cm, clip]{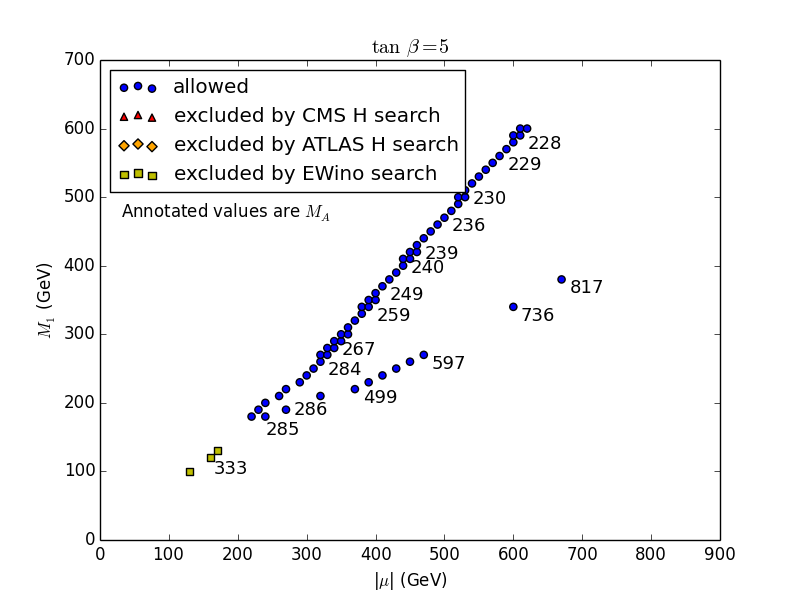}
	\includegraphics[width = 8.16cm, clip]{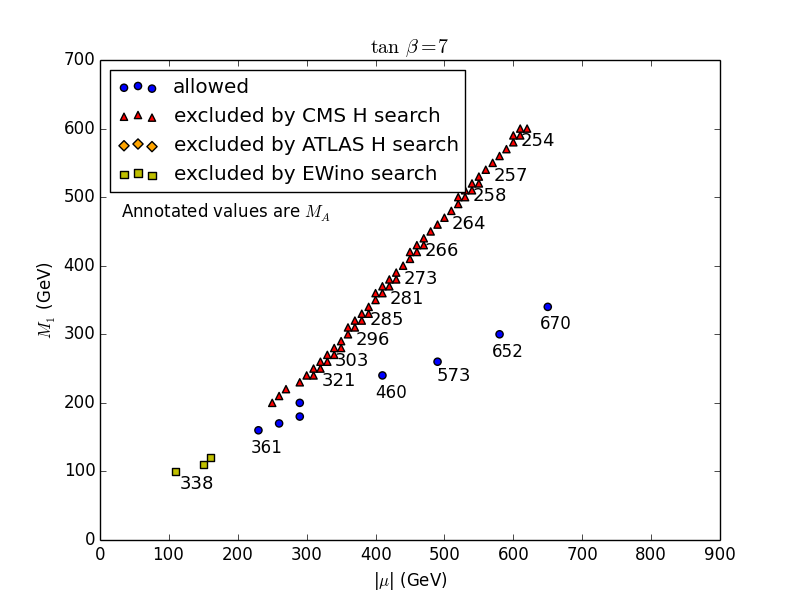}
	\includegraphics[width = 8.16cm,clip]{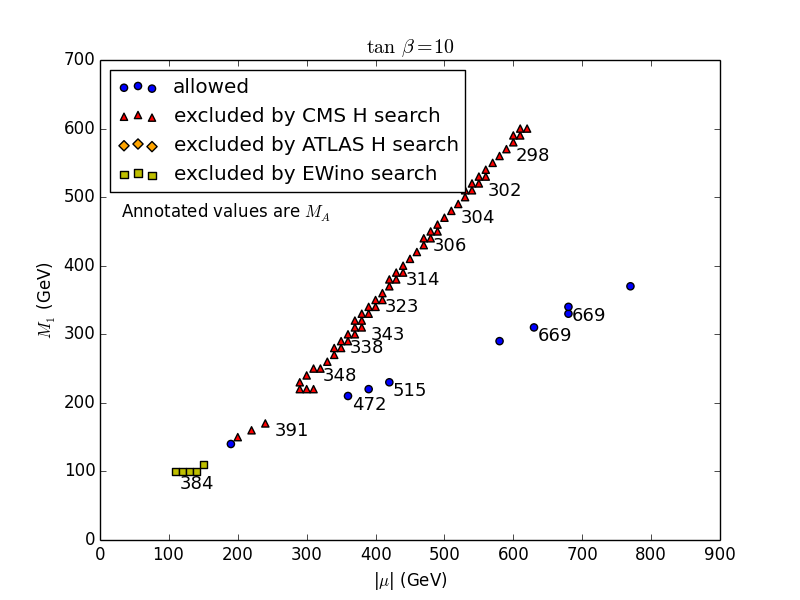}
	\includegraphics[width = 8.16cm,clip]{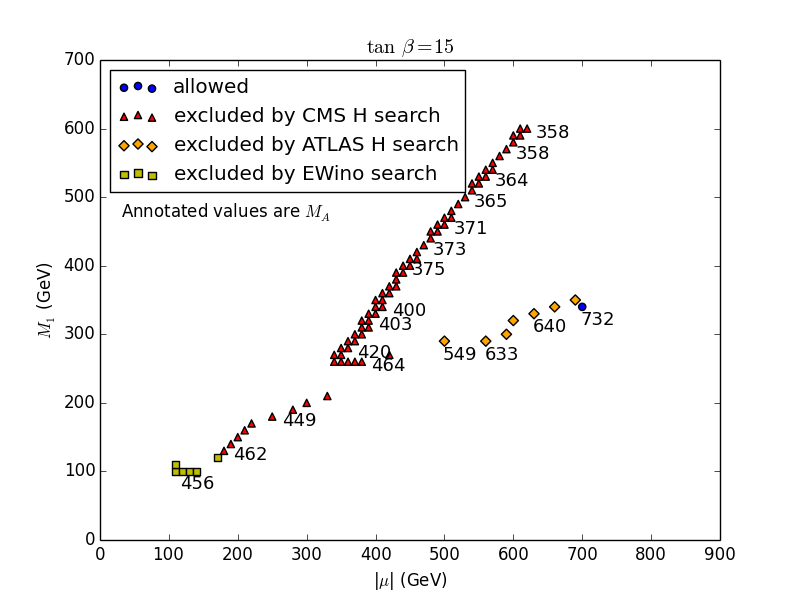}
	\caption{Plot analogous to Fig. \!\ref{fig:MA-muM1} with $M_A$ chosen at the upper limit consistent with  $\sigma_p^{SI} = 10^{-11}$~pb. Only exclusions from the CMS $\phi \rightarrow \tau \tau$ search are shown.}
	\label{fig:MA-muM1R}}
\end{figure}

We also investigate the effect of choosing $M_A$ at the lower or upper limit consistent with  $\sigma_p^{SI} = 10^{-11}$~pb on the exclusion status of our model.  We survey the $|\mu|-M_A$ plane twice more, choosing $M_A$ to be at the lower and upper limits. These new data are plotted in Fig. \!\ref{fig:MA-muM1L} and Fig. \!\ref{fig:MA-muM1R}.

From Fig. \!\ref{fig:MA-muM1L} and Fig. \!\ref{fig:MA-muM1R} we see that, in the well-tempered region, the exclusion bounds are largely independent of which $M_A$ we choose to achieve the  $\sigma_p^{SI} \leq 10^{-11}$~pb condition. 
This stems from the fact that, in the well-tempered region, the variation of $M_A$ from the values consistent with the blind spot scenario to the values leading to $\sigma_p^{SI} = 10^{-11}$~pb is 
$\Delta M_A \simeq 10-20$~GeV. Hence, the well-tempered region is still excluded for $\tan \beta \geq 7$. On the other hand, in the A-funnel region $\Delta M_A \simeq 100 - 300$~GeV for comparable points in Fig. \!\ref{fig:MA-muM1L} and Fig. \!\ref{fig:MA-muM1R}, resulting in a greater disparity between exclusions in the $|\mu|-M_1$ plane, as is especially evident for the $\tan \beta = 15$ plots.

\subsection{Electroweakino Search}

\begin{figure}[tbh]{
	\includegraphics[scale=0.6]{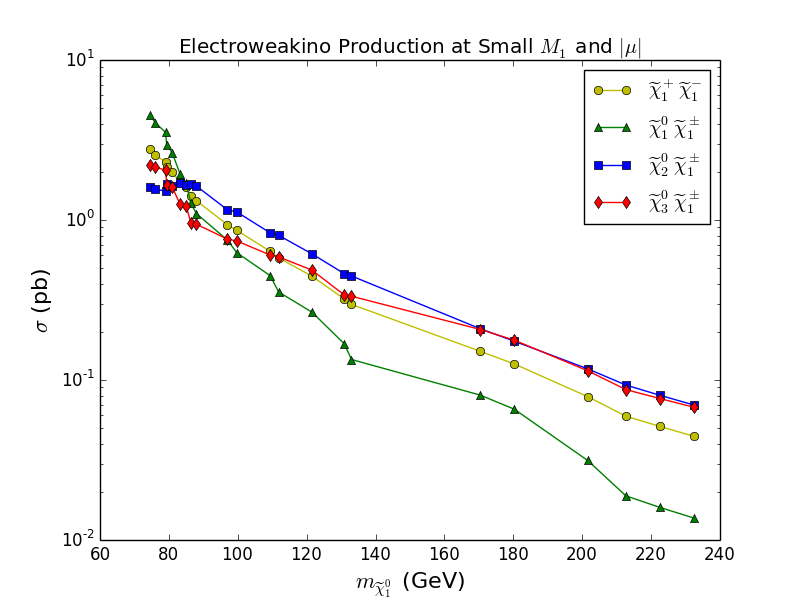}
	\caption{Leading order cross sections for various electroweakino pair productions. Each data point corresponds to a blind spot with small $M_1$ and $|\mu|$ in the well-tempered region ($\mu \simeq M_1$) that was not excluded by the $\phi \rightarrow \tau \tau$ searches, although one outlier is not shown for the sake of readability. The plot incorporates points from all values of $\tan \beta$, since $\tan \beta$ does not have a significant effect on $\sigma$. For these data points $M_1 \simeq |\mu|$, and changes in the sign of $|\mu|-M_1$ can occur between data points. Thus the composition of the electroweakinos changes between data points as well. This results in bumps in the depicted curves since bino-like electroweakinos have lower cross sections than Higgsino-like electroweakinos of comparable mass.}
	\label{fig:EWinoxsection}}
\end{figure}

The $\phi \rightarrow \tau \tau$ searches leave open a small region of parameter space where $M_A$ is sufficiently large to avoid 
these constraints, with  $M_1, |\mu| < 200$ GeV.  The neutralinos and charginos are light in this region, so electroweakino searches at LHC become relevant. The most stringent constraints are obtained from studying the decay products of  the associated production of charginos and neutralinos, $\widetilde\chi_2^0$$\widetilde\chi_1^\pm$. Since our slepton masses have been set high,  the branching ratio for the decay of  the second lightest neutralino and the lightest chargino $\widetilde \chi_2^0$, $\widetilde \chi_1^{\pm}$, into sleptons is negligible. In addition, since $m_{\widetilde \chi_2^0} \simeq m_{\widetilde \chi_1^{\pm}} < M_A + m_\no$, the decay of $\widetilde \chi_2^0$ , $\widetilde \chi_1^\pm$ into heavy Higgs bosons is also negligible. This leaves $ \widetilde \chi_2^0 \widetilde \chi_1^\pm \rightarrow WZ \widetilde \chi_1^0 \widetilde \chi_1^0$ and $\widetilde \chi_2^0 \widetilde \chi_1^\pm \rightarrow Wh \widetilde \chi_1^0 \widetilde \chi_1^0$ as the only viable decay channels, with the final state containing a $Z$ being the most sensitive one. In addition to the decay of $\widetilde \chi_2^0\widetilde \chi_1^\pm$, the decay of $\widetilde \chi_3^0 \widetilde \chi_1^\pm$ is also a significant contributor to the $WZ \widetilde \chi_1^0\widetilde \chi_1^0$ final state. Production cross sections, computed with Prospino2~\cite{prospino}, for these and other electroweakino pairs are shown in Fig.~\ref{fig:EWinoxsection}. In the region of small $M_1$ and $|\mu|$, the masses of $\widetilde \chi_2^0$ and $\widetilde \chi_3^0$ are close, and thus the decays $\widetilde \chi_2^0 \widetilde \chi_1^\pm \rightarrow WZ \widetilde \chi_1^0 \widetilde \chi_1^0$ and $\widetilde \chi_3^0 \widetilde \chi_1^\pm \rightarrow WZ \widetilde \chi_1^0 \widetilde \chi_1^0$ are difficult to distinguish experimentally.  Assuming this final state, CMS excludes a bounded region in the $m_{\widetilde \chi^0_1}-m_{\widetilde \chi^2_0}$ plane \cite{CMS:EWino2016}. We exclude data points from the model according to the CMS bounds. To be more conservative in excluding points from the model, we use the mass $m_{\widetilde \chi_3^0}$ instead of $m_{\widetilde \chi_2^0}$ when testing points against the bounds in Ref.~\cite{CMS:EWino2016} A caveat to this method is that the CMS bounds assume wino-like $\widetilde \chi_2^0$ and $\widetilde \chi_1^\pm$, whereas for our data these electroweakinos are higgsino-like. The production cross sections for wino-like $\widetilde \chi_2^0\widetilde\chi_1^\pm$ are typically four times larger than those for higgsino-like electroweakinos \cite{CERN:EWxsec}. Even considering there are two higgsinos, the total higgsino production cross section is about half the wino production cross section, so the true bounds on our data are weaker than those presented in \cite{CMS:EWino2016}\footnote{For a recent analysis, see Ref~\cite{Han:2016qtc}}. We find that these electroweakino searches do constrain this region, as shown by the yellow points in Figs.~\ref{fig:MA-muM1}-\ref{fig:MA-muM1R}. Although the displayed bounds are generous for the Higgsino-like electroweakinos in our data, we will show in subsection D that this region of parameters is also excluded by recent IceCube results~\cite{cube}. 

The High Luminosity-LHC would extend the scope of the electroweakino searches, and could probe up to 150 GeV to 600 GeV depending on the mass of the LSP~\cite{CMS:2013xfa,Han:2016qtc}. A 100 TeV collider can further extend the reach. For instance, when $m_{\no}$ is below 500 GeV, a 100 TeV collider with 3 ab$^{-1}$ can make a discovery of a higgsino on the order of 1.5 TeV in the trilepton channel~\cite{Gori:2014oua}. In addition, in the well-tempered region, where $M_1\sim |\mu|$, a future 100 TeV collider with 3 ab$^{-1}$ can be sensitive to a mixed bino-higgsino LSP up to 1 TeV, and can reach 5$\sigma$ discovery for $m_{\no} \lesssim$ 165~-~420~ GeV depending on the mass difference between the two lightest neutralinos and the treatment of systematics~\cite{Low:2014cba}.


\subsection{Precision Higgs Measurements}
\label{sec:smhiggs}
In the well-tempered region, as $M_A$ is light, there could be some tension with the precision Higgs data. At the tree level, the 125 GeV Higgs coupling to bottom-quarks in the MSSM is given by~\cite{Gunion:1984yn}
\begin{equation}
\frac{g_{hbb}}{g_{hbb}^{SM}} = -\frac{\sin\alpha}{\cos\beta} = \sin(\beta - \alpha) - \tan\beta\cos(\beta-\alpha). 
\end{equation}
The first term in the left-hand side of this expression, $\sin(\beta-\alpha)$, gives the ratio of the coupling of the Higgs to weak vector bosons
to its SM value. In order to reproduce the proper Higgs phenomenology, it should be close to one. Therefore, the corrections
to the bottom coupling are controlled by the second term. One can work out an approximate expression for the value of this correction
in the MSSM~\cite{Carena:2014nza}, at the one-loop level, namely
\begin{equation} \label{tbcbma}
\tan\beta \; \cos(\beta-\alpha)\simeq \frac{-1}{m_H^2-m_h^2}\left[m_h^2-m_Z^2 \cos(2\beta)+
\frac{3m_t^4 X_t(Y_t-X_t)}{4\pi^2 v^2
  M_S^2}\left(1-\frac{X_t^2}{6M_S^2}\right)\right]\,.
\end{equation}
In the above, $M_S$ is the average stop mass, $X_t = A_t - \mu/\tan\beta$, $Y_t = A_t + \mu \tan\beta$, $A_t$ is the trilinear Higgs stop coupling and $\mu$ is the Higgsino
mass parameter.  The last term denotes the one-loop radiative corrections induced by the interaction of the Higgs bosons with the
third generation squarks. At sizable values of  $\tan\beta$  we can rewrite the above expression in the following approximate form,
\begin{equation} \label{tanbcbma}
\tan\beta\; \cos(\beta-\alpha)\simeq \frac{-1}{m_H^2-m_h^2}\left[m_h^2+m_Z^2+
\frac{3m_t^4 }{4\pi^2 v^2
  M_S^2}A_t\mu \tan\beta\left(1-\frac{A_t^2}{6M_S^2}\right)\right]\,.
\end{equation}
Since in order to obtain the proper Higgs mass in the MSSM the stop masses should be of the order of 1~TeV~\cite{Draper:2013oza},\cite{Vega:2015fna},\cite{Lee:2015uza},  and the value of $A_t < 3 M_S$
due to vacuum stability constraints~\cite{Blinov:2013fta}, it is clear that for the values of $\mu$ and $\tan\beta$ under consideration, the radiative corrections give only a small 
correction and the deviations of the bottom coupling from its SM value are well characterized by the first two  terms inside the square bracket on the right-hand side of Eq.~(\ref{tanbcbma}). 

For instance, 
in the well-tempered region, when $\tan\beta$ is about 5, and the lightest neutralino is about 600~GeV, $M_A$ is about 220~GeV as shown in Fig.\!~\ref{fig:MA-muM1}. This leads to a bottom coupling that is about 70$\%$ higher than the SM value, which is about 4$\sigma$ above the current central value and therefore ruled out by current Higgs precision measurements~\cite{Khachatryan:2016vau,ATLAS:Htautau}.  We stress that this enhancement in the bottom coupling would lead to a large enhancement of the total width and therefore a suppression of the branching ratios of all other decay channels. In the region where $M_A$ is larger, as approaching the decoupling limit, this tension is eased. For example, when $\tan\beta$ is about 5 and $M_A$ is about 350~GeV, the bottom coupling to the Higgs is only about 20$\%$ higher than the SM, which corresponds to about two standard deviation of the experimental result (the current fit to the bottom Yukawa coupling shows a suppression of it~\cite{Khachatryan:2016vau}\footnote{The current fit also indicates an enhancement of the top Yukawa coupling. A suppression (enhancement) of the bottom (top) Yukawa coupling is difficult to achieve in the MSSM, but possible in the NMSSM~\cite{Badziak:2016exn,Badziak:2016tzl}.}).  
Therefore, values of $M_A$ larger than about 350~GeV are necessary to be in agreement with precision electroweak data~\cite{ATLAS:Htautau}. 
Such large values of $M_A$, however, are not consistent with the blind spot in the well-tempered region and therefore this MSSM scenario leads to tension with precision Higgs measurements. 

The tension may be relaxed in two ways. Within the MSSM, one can consider the possibility of a SI DDMD cross section 
larger than the $\sigma_{SI} \simeq 10^{-11}$~pb assumed in this scenario.  For instance, if we instead consider the maximal cross sections allowed by the current LUX bounds, as shown in Fig.~\ref{fig:MAubLUX}, then the well-tempered region can coexist with the precision Higgs data for neutralino masses of order 300~GeV and heavy CP-even Higgs masses $m_H > 350$~GeV.

On the other hand, one can consider the possibility of extending the MSSM. In the simplest of such extensions, the NMSSM~\cite{Ellwanger:2009dp}, in which
a singlet superfield is added to the spectrum, the couplings and mixing of the CP-even Higgs bosons are modified by the appearance of
new couplings and mixing with the singlet CP-even Higgs. In particular, the superpotential coupling $\lambda$ of the singlet superfield
to the Higgs doublets plays a significant role in defining the corrections to the bottom coupling to the lightest CP-even Higgs. 
If one considers the limit in which the singlet sector masses are raised by supersymmetry breaking terms, then the neutralino and Higgs
boson particles in the low energy theory become identical as those ones in the MSSM. Moreover, one can consider the effective
2x2 CP-even Higgs mixing mass matrix after decoupling of the singlet fields, from where one can demonstrate that the 
bottom coupling has the same expression as in the MSSM, but the value of $\tan\beta \cos(\beta-\alpha)$ is now given by~\cite{Carena:2015moc}
\begin{equation} \label{tbcbma1}
\tan\beta \; \cos(\beta-\alpha)\simeq \frac{-1}{m_H^2-m_h^2}\left[m_h^2+m_Z^2 -  \lambda^2  v^2 +
\frac{3m_t^4 }{4\pi^2 v^2
  M_S^2}A_t\mu \tan\beta\left(1-\frac{A_t^2}{6M_S^2}\right)\right]\,.
\end{equation}
From Eq.~(\ref{tbcbma1}) it follows that for moderate or large values of $\tan\beta$, values of the coupling $\lambda$ in the range $\lambda \simeq 0.6$--$0.7$ lead to small deviations of the bottom coupling with respect to the SM value, even for values of $m_H \simeq 200$~GeV.  
Hence, in the NMSSM, for heavy singlets and singlinos, the well-tempered region can be brought to agreement with precision Higgs measurements.  

\subsection{Spin-Dependent and Indirect Detection Constraints}

\begin{figure}[tbh]{
	\includegraphics[scale=0.6]{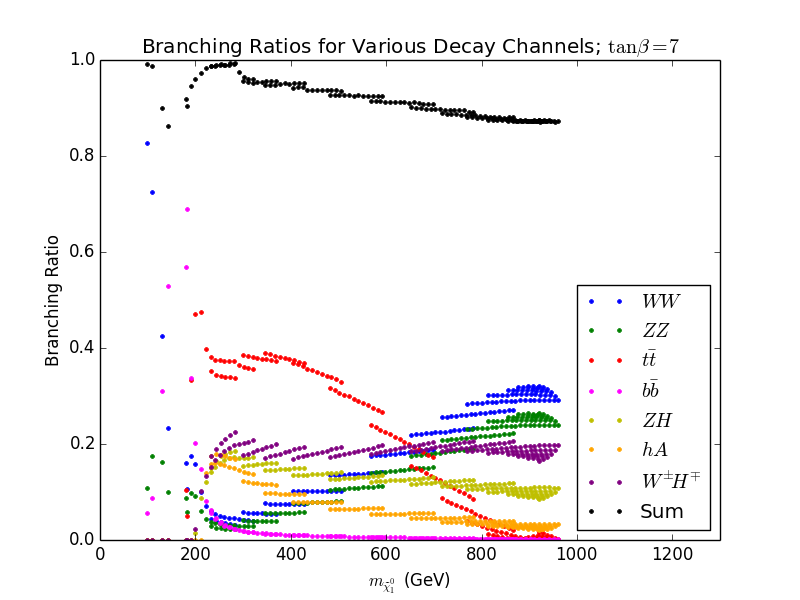}
	\caption{Branching ratios for dark matter annihilation products, including $W^+W^-$ (blue), $ZZ$ (green), $t\bar t$ (red), $b\bar b$ (pink), $ZH$ (yellow), $hA$ (orange), $W^\pm H^\mp$ (purple), and their sum (black). We see that for $m_\no < 200$~GeV, the $W^+W^-$, $ZZ$, and $b \bar b$ decay channels dominates. At $m_\no \simeq 200$~GeV, the $t \bar t$ decay channel becomes prevalent, but begins to diminish for large $m_\no$. For $m_\no > 200$~GeV, the branching ratios for annihilation into $W^+W^-$ and $ZZ$ are similar, and the decays into $ZH, hA,$ and $W^\pm H^\mp$ become significant.}
	\label{fig:icecubeBR}}
\end{figure}

Many direct and indirect detection experiments have also placed constraints on a DM particle interacting with a nucleus via spin dependent (SD) scattering. Spin dependent direct detection searches have been performed by the LUX~\cite{luxfinal} and XENON100~\cite{xenon100} experiments, but PICO~\cite{pico} outperforms
both of them. These exprimental results, however, do not set a strong bound on the regions of parameters explored in this paper.  The exception 
comes from
IceCube, which considers the detection of neutrinos coming from Dark Matter trapped and annihilating in the sun. The limits depend strongly
on the annihilation channel but become strong for neutralinos annihilating into pairs of $W^+W^-$ or $ZZ$ gauge bosons~\cite{cube}, for which the current cross section limits 
are just below $\sigma_p^{SD} \simeq 10^{-4}$ pb. In our model, this annihilation channel is significant in  the well-tempered neutralino region, where the higgsino component of the neutralino is largest and the Z boson coupling required for Z boson exchange in SD scattering is not suppressed. However, for $130~\text{GeV} \lesssim m_\no \lesssim 190$~GeV, the LSP also annihilates significantly  into $b \bar b$, for which the IceCube bounds are weak. Furthermore, above $200$~GeV the LSP begins annihilating into top quarks, and the branching ratio into $W$ bosons diminishes. The remaining branching ratio is predominantly in the decay channels $ZH, W^\pm H^\mp$, and $hA$, which are not analyzed by IceCube. Assuming them to have bounds similar to the $t\bar t$ bounds, they will modify the constraints only slightly. Fig.~\ref{fig:icecubeBR} displays the branching ratios for these channels in the well-tempered region.

\begin{figure}[tbh]{
	\includegraphics[scale=0.6]{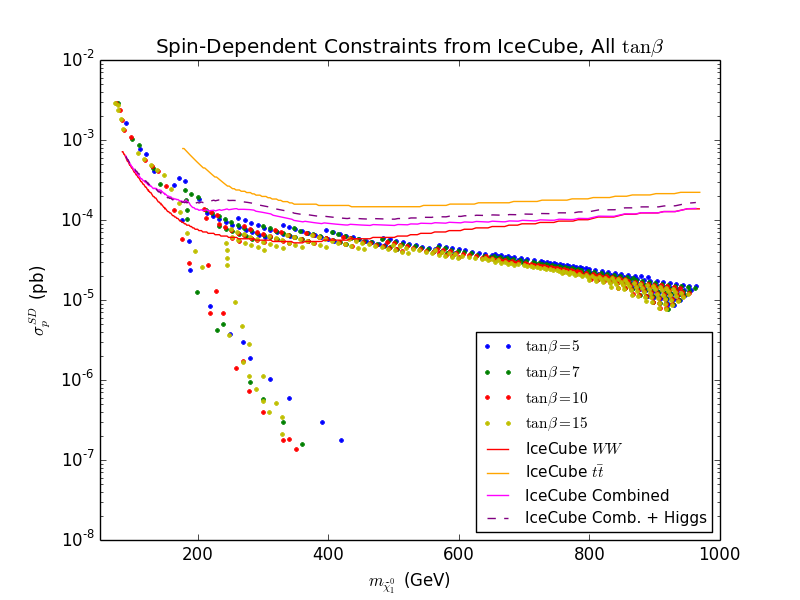}
	\caption{Calculated spin dependent cross sections for $\tan\beta = 15$  (yellow dots), $\tan\beta = 10$ (red dots), $\tan\beta = 7$ (green dots)
	and $\tan\beta = 5$ (blue dots). The upper branch correspond to the well-tempered region, while the lower one corresponds to the A-funnel region.  The red solid line represents the current 90\% C.L. bound on the Spin dependent cross section coming from IceCube for annihilation into $WW$, and the orange line for annihilation into $t \bar t$. The bound for annihilation in to $ZZ$ is very similar to the bound for annihilation into $WW$. The magenta and dashed purple lines combine these bounds, weighting them by the branching ratios for our data, with the dashed purple line further taking into account the decays into $ZH, W^\pm H^\mp$, and $hA$. The bounds for decay in $b \bar b$ are several orders of magnitude weaker and are not shown.}
	\label{fig:icecubeall}}
\end{figure}

Figure~\ref{fig:icecubeall} show the current 90\% confidence level bounds on the blind spot scenarios coming from IceCube for the different values of $\tan\beta$ analyzed in this article. The lines labeled ``IceCube combined'' are the square roots of the harmonic means of $\sigma_i^2$ weighted by the branching ratios for our data, where $i$ runs across the various decay channels. For the solid magenta line we have considered only the decays into $W^+W^-, ZZ$, and $tt$, rescaling the branching ratios to sum to one. In the dashed purple line we have included the decays $ZH, W^\pm H^\mp$, and $hA$, approximating them to make up the remainder of the branching ratio and have bounds on the similar to the $t \bar t$ bounds from IceCube. The latter method excludes dark matter masses in the well tempered region for $m_\no \lesssim 200$~GeV, while the former method has a slightly stronger bound, excluding masses less than 210~GeV.

Finally, let us comment on indirect Dark Matter detection constraints. While the constraints from the Fermi experiment~\cite{Ackermann:2015zua} do not affect the scenarios discussed in this article, there has been recent analyses of the AMS antiproton flux data~\cite{Aguilar:2016kjl} that
claim strong constraints on thermal dark matter annihilating into bottom-quark pairs, with masses between 150~GeV and     450~GeV~\cite{Cuoco:2016eej},\cite{Cui:2016ppb}.  Although there are large  uncertainties having to do with propagation, solar modulation and antiproton production cross sections, if these bounds hold, the A-funnel region will be constrained to values of $\mu$ and $M_A$ larger than about 1~TeV.

\section{Conclusions}
\label{sec:conc}

In this article we have studied the constraints and future probes of Dark Matter in the MSSM, in the case in which all scalar leptons and 
quarks are heavy.  In particular, we have considered scenarios within the MSSM, in which the SI DDMD cross section
is suppressed due to destructive interference between the light and heavy CP-even Higgs exchange amplitudes. We have shown that the
proper relic density may be obtained in both the well-tempered neutralino region as well as in the A-funnel region. In the well-tempered
region, the values of the heavy Higgs boson masses are lower than twice the top quark mass and this region of parameters may be
efficiently probed by searches for production of heavy Higgs bosons decaying into $\tau$-lepton pairs. Current searches already
restrict the value of $\tan\beta < 7$ in this region of parameters and future searches can probe the whole region consistent with the
blind spot scenario. Moreover, the IceCube are in tension with the well-tempered scenario for neutralino masses lower than
200 GeV.  Furthermore, for neutralino masses larger than about 400~GeV, allowed values of the CP-odd Higgs mass may be in tension with those required to get consistency with precision Higgs measurements and the realization of this scenario may require a Higgs sector that goes beyond the MSSM one, like the one that is obtained in the NMSSM for heavy scalar and fermion singlets. Current bounds, however, allow the realization
of the well-tempered scenario for neutralino masses of the order of 300 GeV, $\tan\beta \simeq 5$  and heavy Higgs bosons of about 400~GeV.

In the A-funnel region, the heavy Higgs boson masses are larger and, therefore, this region of parameters cannot be
fully probed by current or future LHC searches for heavy Higgs bosons. This is particularly true at lower values of $\tan\beta$, where the
decay into top-quark pairs tends to be comparable or larger than the one into $\tau$-lepton pairs.  On the other hand, electroweakino
searches present an alternative way of probing this scenario, although it is efficient only for low values of $\mu$ and $M_1$.
Finally, for values of the neutralino mass lower than 450~GeV, this region of parameters may be in tension with recent AMS antiproton
data, what may require to go to values of $M_A$ and $\mu$ larger than about 1~TeV.

\section*{Acknowledgements}
 Work at the University of Chicago is supported in part by U.S. Department
of Energy grant number DE-FG02-13ER41958.
Work at ANL is supported in part by the U.S. Department of Energy under 
Contract No. DE-AC02-06CH11357.  P.H. is partially supported by the US. 
Department of Energy under Contract No. DE-FG02-13ER42020, and thanks 
the Mitchell Institute for Fundamental Physics and Astronomy for support. We would like to thank M. Carena, 
F. Ferrer, D. Hooper, N.R. Shah, and T. Tait for useful discussions and comments.

\bibliographystyle{utphys}
\bibliography{newbsrefs.bib}
\end{document}